\documentclass[%
 reprint,
 amsmath,amssymb,
 aps,
prb,
]{revtex4-2}

\usepackage{textgreek}

\usepackage{float}
\usepackage{bm}
\usepackage{hyperref}

\usepackage{graphicx}
\usepackage{dcolumn}
\usepackage{bm}

\usepackage{siunitx}
\usepackage{xcolor}

\newcommand{\supplementarysection}{%
  \setcounter{figure}{0}
  \let\oldthefigure\thefigure
  \renewcommand{\thefigure}{S\oldthefigure}
}

\begin{document}

\preprint{APS/123-QED}

\title{Tracking phase synchronization between flagella in the time-frequency domain resolves photophobic response}

\author{Lucas FEDERSPIEL, Françoise ARGOUL, Antoine ALLARD}
\email{francoise.argoul@u-bordeaux.fr and antoine.allard@u-bordeaux.fr}
\affiliation{Univ. Bordeaux, CNRS, LOMA, UMR 5798, F-33400 Talence, France}%

\author{Jorge ARRIETA}
\affiliation{Departmento de Física, Universitat de les Illes Balears, 07122, Palma, Spain}

\author{Marco POLIN}
\affiliation{Mediterranean Institute for Advanced Studies, IMEDEA, UIB-CSIC, Esporles, 07190, Spain}

\date{\today}

\begin{abstract}

The unicellular microalga \textit{Chlamydomonas reinhardtii} (CR) is well known for its bi-flagellated swimming in response to light stimuli.
This work aims to study the resynchronization of CR flagella after a high light intensity stimulus, known as photoshock.
The synchronization is estimated thanks to a quantity defined as the Phase Synchronization Index (PSI). The originality of this approach is to perform a time-frequency computation of a complex PSI based on continuous wavelet transform.
Thanks to this analysis, we distinguish three swimming stages involving different frequency bands and phase shifts: synchronized breaststroke swimming, undulatory backward swimming, and resynchronization.
This approach also reveals the presence of signal harmonics that set the photoshock response, independently of cell variability. 
Our results suggest that CR modulates the balance between spectral beating modes, providing a mechanism for robust adaptation to sudden environmental stresses.

\end{abstract}


\maketitle

\twocolumngrid 
\newpage

\section{Introduction}

Oscillators are systems that exhibit repetitive behavior around a steady state. They are found across a wide range of fields, including mechanics (\textit{e.g.}, pendulums and vibrating strings), optics (\textit{e.g.}, electromagnetic oscillator), or active matter systems both inanimate and living \cite{riedl_synchronization_2023}. 
In biology, rhythmic processes such as neuron spikes \cite{izhikevich_dynamical_2006}, cardio-respiratory activity \cite{schafer_synchronization_1999}, and circadian rhythms \cite{schibler_steven_2024}, display oscillatory dynamics across multiple spatial and temporal scales. 
While the behavior of one isolated oscillator is generally well described, two or more coupled oscillators exhibit much more complex responses \cite{strogatz_coupled_1993,denis_unclocklike_2024}. 
The concept of synchronization, where oscillators adjust their dynamics to cooperate, was first introduced in physics by Huygens in the 17th century, who observed phase or anti-phase coupling of two pendulum clocks \cite{huygens_horologium_1673}. 

In living systems, synchronization is essential for robust physiological function: it coordinates gait during during locomotion, supports coherent brain activity, and drives fluid transport by ciliary carpets \cite{pikovsky_synchronization_2002,button_periciliary_2012,hu_multiflagellate_2024}. 
However, many biological oscillators exhibit nonstationary and nonlinear behavior. This means that their natural frequency can change dynamically in response to internal processes or environmental perturbations \cite{holland_control_1997,guillet_uncertainty_2024}. 
In such systems, transient coupling between oscillators appears, making synchronization detection challenging. 
Addressing this challenge requires time-resolved tools that track not only the instantaneous frequency evolution of each oscillator, but also their relative phase dynamics, which govern synchronization.

Here, we introduce a time-frequency synchronization framework, based on continuous wavelet transform (CWT) capable of capturing transient and frequency-dependent coupling, enabling insight into nonstationary biological oscillators \cite{torresani_analyse_1995,torrence_practical_1998,kronland-martinet_analysis_1987,schack_quantification_2005,liang_synchronization_2016,grubov_harnessing_2025}.
Within this framework, we introduce a time- and frequency-resolved Phase Synchronization Index (PSI), that quantifies transient coupling and reveals how synchronization evolves across multiple time scales.
This method not only detects when synchronization is present, but also identifies how characteristic frequency bands shift during external perturbations. 
We apply this framework to experimental data on flagella synchronization in the green alga \textit{Chlamydomonas reinhardtii} (CR), and complement these results with a minimal phase-coupling model with time-varying intrinsic frequencies \cite{adler_study_1946,kuramoto_chemical_1984}. The later allows us to probe how varying coupling strength can help interpret the observed experimental behaviors.

CR is a model organism to study motility, phototaxis or sensory responses \cite{harris_chlamydomonas_1989}.
CR typically swims using a a breaststroke-like motion, in which its two flagella beat synchronously \cite{liu_transitions_2018}.
This coordination results from intracellular coupling, which is necessary \cite{wan_coordinated_2016} and sufficient \cite{zorrilla_role_2025} for the wild-type strain CC125. In the case of the CR strain \textit{ptx1}, known to have a weaker intracellular coupling, hydrodynamic coupling impacts resynchronization \cite{friedrich_hydrodynamic_2016,zorrilla_role_2025}.
The ability of CR to thrive in various conditions makes it an ideal organism for exploring adaptive responses to changes in environmental cues \cite{goldstein_noise_2009,hegemann_light-induced_1989,ruffer_flagellar_1991,arrieta_phototaxis_2017,ramamonjy_nonlinear_2022,zorrilla_role_2025}.
For instance, CR senses light through rhodopsin-type photoreceptors (channelrhodopsins) in its eyespot, with peak sensitivity in the blue-green range ($\sim\qtyrange[range-units=single,range-phrase=-]{450}{500}{\nano\meter}$) \cite{berthold_channelrhodopsin-1_2008,jeanneret_brief_2016}. These photoreceptors trigger photocurrents that modulate flagellar beating and drive both phototactic and photophobic behaviors \cite{ruffer_flagellar_1991,ruffer_flagellar_1998}.
Under strong light stimulation, such as brief (around \qty{50}{\milli\second}) and high irradiance ( $>\qtyrange[range-units=single,range-phrase=-]{10}{100}{\watt\per\meter\squared}$) blue-green pulses, CR undergoes a photoshock response, transiently reversing its swimming direction through a switch to a high-frequency, undulatory beating mode \cite{ruffer_flagellar_1998,holland_control_1997,ueki_dynein-mediated_2018}.
Although steady-state flagellar coordination in CR has been extensively studied, little is known about how synchronization dynamically breaks and re-emerges following such perturbations.

In this work, we address this gap by studying flagellar resynchronization during photoshock. 
Since the oscillatory motion of flagella is performed at low Reynolds number, we infer oscillatory dynamics from the surrounding flow field instead of directly tracking flagellar shapes, providing robust estimates of beating frequency and phase \cite{brennen_fluid_1977,boggon_embodied_2025}.
Using our wavelet-based PSI, we resolve the time-dependent synchronization of the two flagella. 
We identify three distinct dynamical stages: (i) a pre-stimulus forward swimming with breaststroke beating around \qty{40}{\hertz}, (ii) a transient high-frequency mode (up to \qty{80}{\hertz}) associated with backward swimming, and (iii) a gradual recovery into the original breaststroke rhythm.
Beyond synchronization, our analysis uncovers the presence and evolution of harmonic components throughout the alga response to photoshock.
Intriguingly, we find that the first harmonic of the breaststroke mode is continuously present, even during photoshock, and becomes dominant in the backward-swimming stage.
The relative energy of these spectral modes shifts gradually, suggesting that CR does not simply switch between discrete beating patterns but rather modulates the balance of coexisting oscillatory modes. 
This continuous interplay between spectral harmonics may underlie the robustness and adaptability of flagellar coordination in changing environments.

\section{Materials and methods}

\begin{figure}[!t]
    \centering
    \includegraphics[width=1\linewidth]{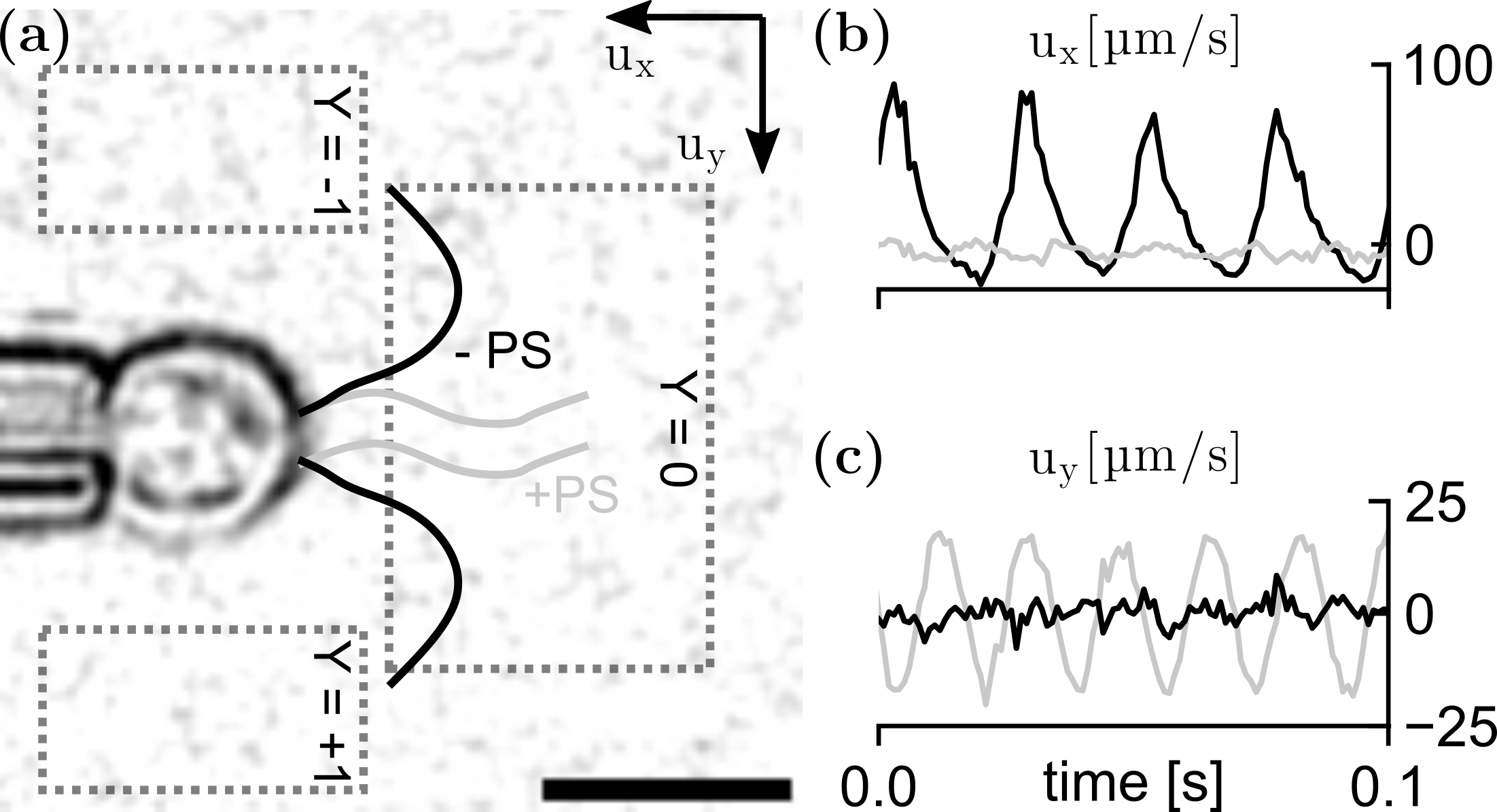}
    \caption{\textbf{Experimental setup}. (a) {\em C. reinhardtii} cell trapped in a micropipette and bathed with tracer beads. Dashed squares indicate the regions of interest where velocity signals are extracted ($Y=0, \pm 1$). Schematic lines represent a screenshot of the flagella before photoshock (black, -PS) and after photoshock (gray, +PS). Scale bar: \qty{10}{\micro\meter}. (b-c) Velocity components along the $x$-axis (b) and $y$-axis (c), averaged in front of the cell ($Y=0$), before photoshock (black) and after photoshock (gray).}
    \label{fig:exp_setup}
\end{figure}

\subsection{Experimental setup}
\textit{Chlamydomonas reinhardtii} strain CC125 (Chlamydomonas Resource Centre, USA) was cultured in liquid medium under continuous agitation at 120 rpm to prevent sedimentation. 
A 14 h light/10 h dark cycle was maintained to synchronize cell division.
Experiments were performed on a bright-field inverted microscope (Nikon Eclipse TE2000-U) modified with a 780 nm LED (Thorlabs, M780L3) for Köhler illumination, a 60X/1.00W water-immersion objective (Olympus), and a high-speed camera (Edgertronic SC1).

Experimental chambers were prepared with \qty{100}{\micro\liter} of algal suspension at a concentration of \num{5e4} cells/mL, bathed with \qty{250}{\nano\meter} diameter polystyrene beads at a volume fraction of 0.37\%. 
Chambers were kept under infrared illumination for at least 30 minutes prior to measurements to stabilize the cells.
Micropipettes were fabricated from borosilicate capillaries (1B100-6, WPI) using a laser puller (Sutter P-2000; parameters: Heat 996, Pull 90, Vel 13, Time 250, Pressure 560), and subsequently forged with a heated filament (MF200-H3, WPI) to refine the tip. 
For each experiment, a pipette filled with water was introduced into the chamber. 
Cells were immobilized by gentle suction (around \qtyrange[range-units=single,range-phrase=-]{10}{100}{\pascal}), and oriented so that both flagella beat within the imaging plane (Fig.~\ref{fig:exp_setup}(a)), though the \textit{cis} and \textit{trans} flagella could not be reliably distinguished experimentally.
Cells remain in place for the all duration of the experiment procedure described below.

\begin{figure}[h!]
    \centering
    \includegraphics[width=1\linewidth]{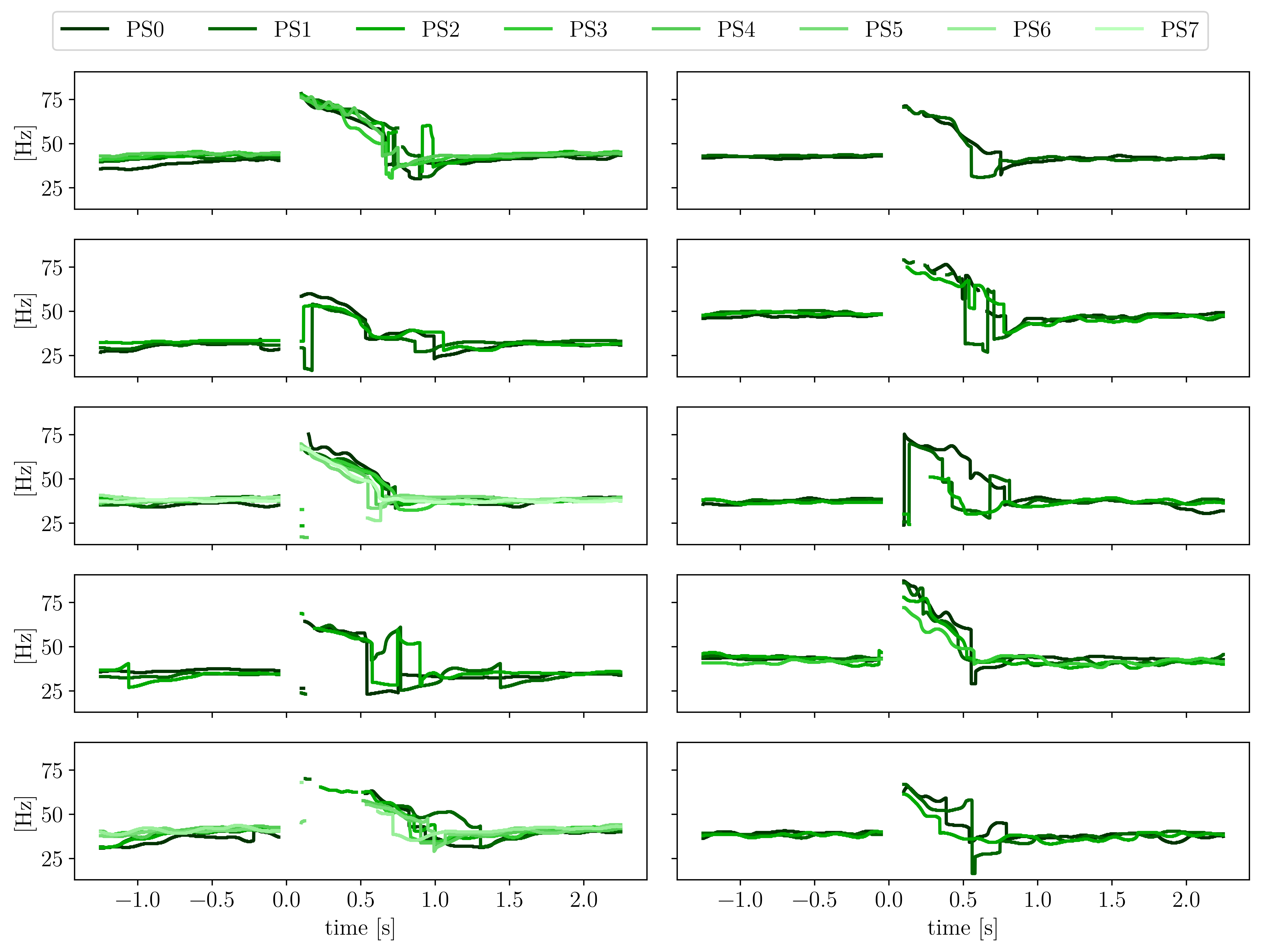}
    \caption{\textbf{CR beating frequency evolution over multiple photoshocks.} The frequencies (green lines) are computed by tracking wavelet modulus maxima on $u_x$ signals in $Y=0$. PS stands for photoshock and the associated number denotes the relative time order of the photoshock. Three algae possess only one photoshock signal, and are therefore not shown in this figure.}
    \label{fig:placeholder}
\end{figure}

Thirteen independent cells were considered. For each, movies of \qty{25}{\second} duration were recorded at 1000 frames per second, with a spatial resolution of $496 \times 496$ px (1 px = \qty{0.222}{\micro\meter}).
Photophobic responses were triggered using a \qty{470}{\nano\meter} LED (Thorlabs, M470L5) mounted laterally to the microscope, reflected by a hot mirror (FM01R, Thorlabs) that illuminate a region larger than the experimental chamber (roughly $\qty{1}{\centi\meter}\times\qty{1}{\centi\meter}$). 
The LED delivered \qty{50}{\milli\second} pulses of high-intensity light ($\sim\qty{10}{\watt\meter\squared}$ irradiance) every \qtyrange[range-units=single,range-phrase=-]{3}{5}{\second}. While the location of the eyespot remains undetermined, the pulse is strong enough to assume that response would be independent of eyespot actual orientation. Each cell was subjected to a maximum of 8 consecutive photoshocks. We did not observe any measurable adaptation of the response over successive stimuli (Supp. Fig.~\ref{fig:placeholder}).

\begin{figure}[h!]
    \centering
    \includegraphics[width=0.6\linewidth]{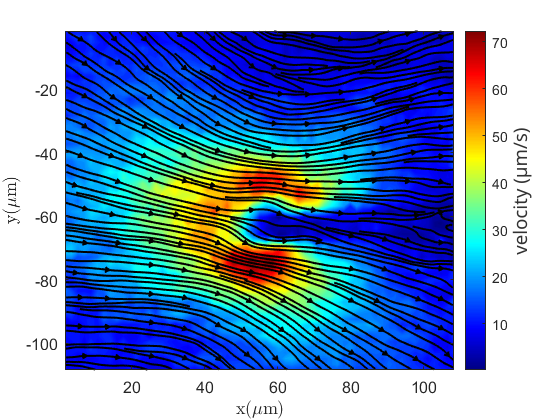}
    \caption{Typical flow field, averaged over multiple beatings and measured using Ghost Particle Velocimetry (GPV).}
    \label{fig:gpv}
\end{figure}

The fluid velocity field was measured using Ghost Particle Velocimetry (GPV) \cite{riccomi_ghost_2018}, with velocity vectors computed over $32 \times 32$ px interrogation windows (\num{17} px overlap) along a rolling window of four frames.
An example of typical flow field is given in Supp. Fig.~\ref{fig:gpv}.
To analyze flagellar dynamics, we extracted velocity signals from three representative regions of interest (Fig.~\ref{fig:exp_setup}(a)). 
Two windows of \qty{20}{\micro\meter}$\times$\qty{10}{\micro\meter}, were located symmetrically on each side of the cell body ($Y=-1$ and $Y=1$), at a distance of approximately \qty{15}{\micro\meter} to predominantly capture the flow generated by the adjacent flagellum. 
A third window of \qty{20}{\micro\meter}$\times$\qty{30}{\micro\meter} was placed in front of the cell ($Y=0$), providing a measure of the bulk swimming flow.
For this central region, the recorded signal reflects the combined contribution of both flagella, without distinction between them.
From each window, velocity signals were decomposed along two orthogonal directions: the $x$-axis, aligned with the main body axis of the cell, and the $y$-axis, perpendicular to it. Fig.~\ref{fig:exp_setup}(b-c) shows representative velocity traces from the bulk ($Y = 0$) before (black) and after (gray) a photoshock.

\subsection{Modeling of coupled oscillators}

Harmonic oscillators can generally be described as $x(t) = A(t) \cos (\varphi(t))$ with $\varphi(t) = \omega t + \varphi_0$, $\varphi(t)$ the phase, $A(t)$ the amplitude and $\omega/2\pi$ the oscillator frequency \cite{friedrich_hydrodynamic_2016}.
For simplicity, we assume that the amplitude $A$ is constant and does not depend on the oscillator, and we concentrate on the coupling dynamics of oscillator phases. 
Note that this assumption implies that either the dynamics of the amplitude are uncoupled to that of the oscillator phase, or the amplitude varies very little, meaning that the system is not much dissipative. 
Such an hypothesis is obviously not fully rigorous in the context of living systems. However, for finite time intervals, dissipation effects could be considered as negligible. 

In a very generic way, phase coupling dynamics can be written as an ordinary differential equation (ODE) system, assuming symmetric coupling:
\begin{equation}
\left\{ 
    \begin{array}{l}
        \dot{\varphi _1} = \omega _ 1 + C(\varphi _1 - \varphi _2) \\
        \dot{\varphi _2} = \omega _ 2 + C(\varphi _2 - \varphi _1)
    \end{array} \; . \label{eq:modeloscillators_1}
\right.
\end{equation}
The coupling function $C(\cdot)$ is $2\pi$ periodic, and in its simplest form, can be written as a single harmonic function: $C(\delta(t)) = B \sin(\delta(t) + {\rm cst})$, with $\delta(t) = \varphi_1(t) - \varphi _2(t)$. We take ${\rm cst}=0$.
The difference of these two equations yields the Adler equation $\dot{\delta} = \Delta \omega + 2B \sin (\delta)$,\cite{adler_study_1946} where $\Delta \omega = \omega_1 - \omega_2$.

\begin{figure}[!t]
    \centering
    \includegraphics[width=1\linewidth]{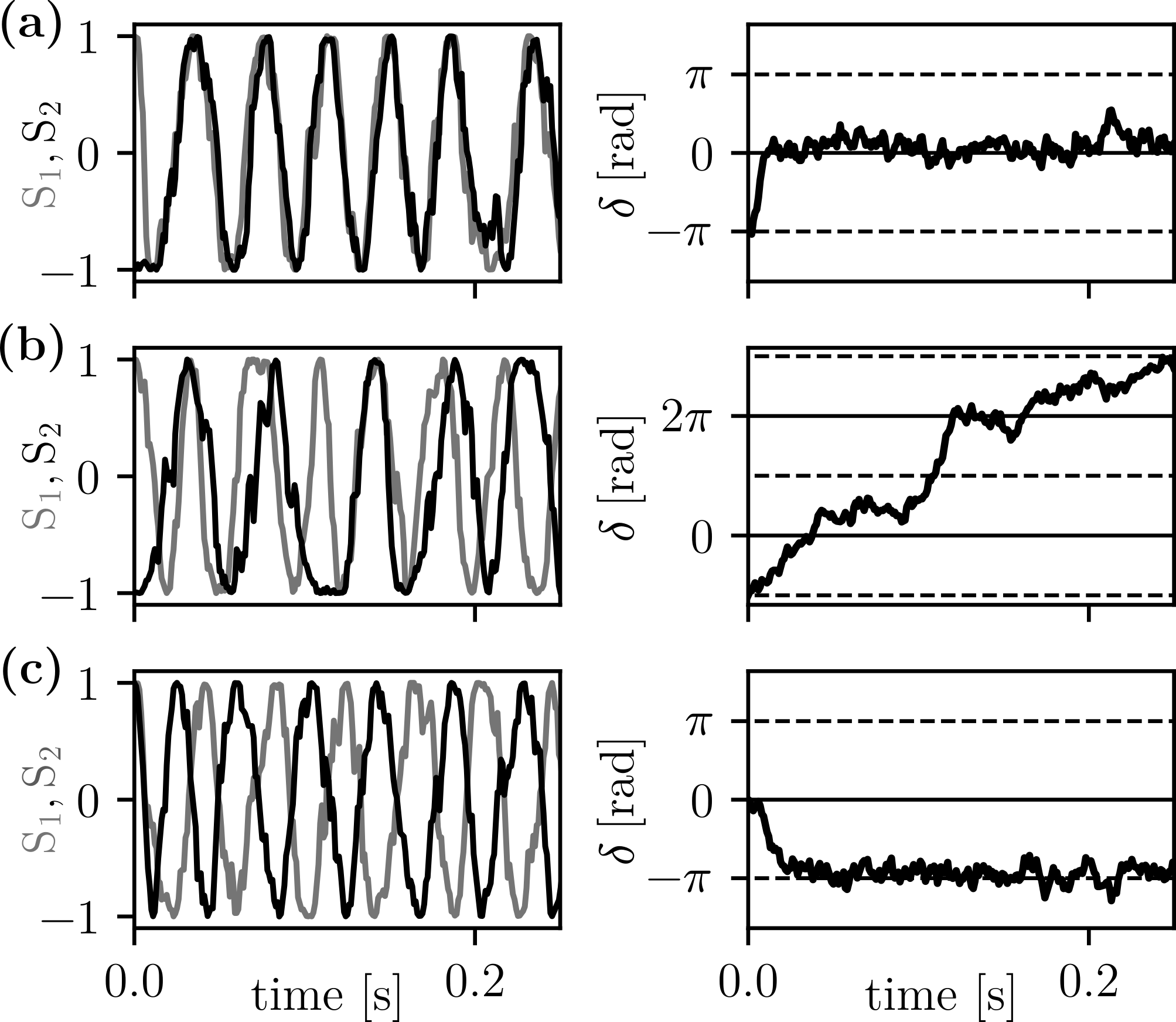}
    \caption{\textbf{Simulations of two coupled oscillators with Gaussian noise, modeled by Eq.~(\ref{eq:S_delta_noise})}.  (a) Strong negative coupling ($B/2\pi=\qty{-20}{\hertz}<0$, $|2B|>\Delta \omega$), with $\varphi_1 (0) = 0$, $\varphi_2 (0) = \pi$: leads to stable in-phase synchronization. (b) Weak negative coupling ($B/2\pi=\qty{-3}{\hertz}<0$, $|2B|<\Delta \omega$), with same initial phases: leads to desynchronization with transient coordination. (c) Strong positive coupling ($B/2\pi=\qty{20}{\hertz}>0$, $|2B|>\Delta \omega$) with $\varphi_1 (0) =\varphi_2 (0) = 0$: leads to stable anti-phase synchronization. Left panels: $S_1=\cos(\varphi_1)$ (gray) and $S_2=\cos(\varphi_2)$ (black). Right panels: $\delta=\varphi_1 - \varphi_2$. Dashed and solid lines represent anti-phase and in-phase locking, respectively.} 
    \label{fig:coupling_sim}
\end{figure}

If we include independent additive Gaussian white noise terms ($\xi_1$ and $\xi_2$) in the phase dynamics, the ODE system (\ref{eq:modeloscillators_1}) becomes a stochastic differential equation (SDE) system:
\begin{equation}
\left\{ 
    \begin{array}{l}
         \dot{\varphi _1}(t) = \omega _ 1 + B \sin (\delta(t)) + \xi _1(t) \\\dot{\varphi _2}(t) = \omega _ 2 - B \sin (\delta(t)) + \xi _2(t) 
    \end{array} \label{eq:S_delta_noise} \; .
\right.
\end{equation}
The noise terms $\xi_1$ and $\xi_2$ follow $\langle \xi_i(t) \rangle = 0$ and $\langle \xi_i(t) \xi_j(t') \rangle = 2D \delta_{ij} \delta(t - t')$, where $\delta_{ij}$ is the Kronecker delta, $\delta(t - t')$ the Delta function and $D$ the phase-diffusion coefficient. To solve this SDE system (\ref{eq:S_delta_noise}), within the Ito formalism, these  coupled SDEs are written in the form of a Wiener process: ${\rm d}\varphi = f(\varphi,t){\rm d}t + G(y,t){\rm d}W(t)$ with $G(y,t)$ the noise coefficients and ${\rm d}W(t)$ independent Wiener increments \cite{rosler_rungekutta_2010}. The numerical computation uses an order 1.0 strong SRK algorithm (SRK scheme SRI2W1 of order (3.0, 1.5)).  For the illustrations of Figs~\ref{fig:coupling_sim} and \ref{fig:PSI_on_sim}, constant noise coefficients were taken, $G_1=G_2=5$  and  $G_1=G_2=1$ respectively.

\subsection{Time frequency analysis of non-stationary signals}
The simplest and most common method for spectral decomposition of signals is the Fourier transform \cite{roddier_distributions_1984}. 
Although effective for stationary signals,  temporal windows must be introduced to restrict the analysis to finite time intervals  when the spectral signature of the signal changes over time as described in \cite{torresani_analyse_1995}.

\begin{figure}[h!]
    \centering
    \includegraphics[width=0.4\linewidth]{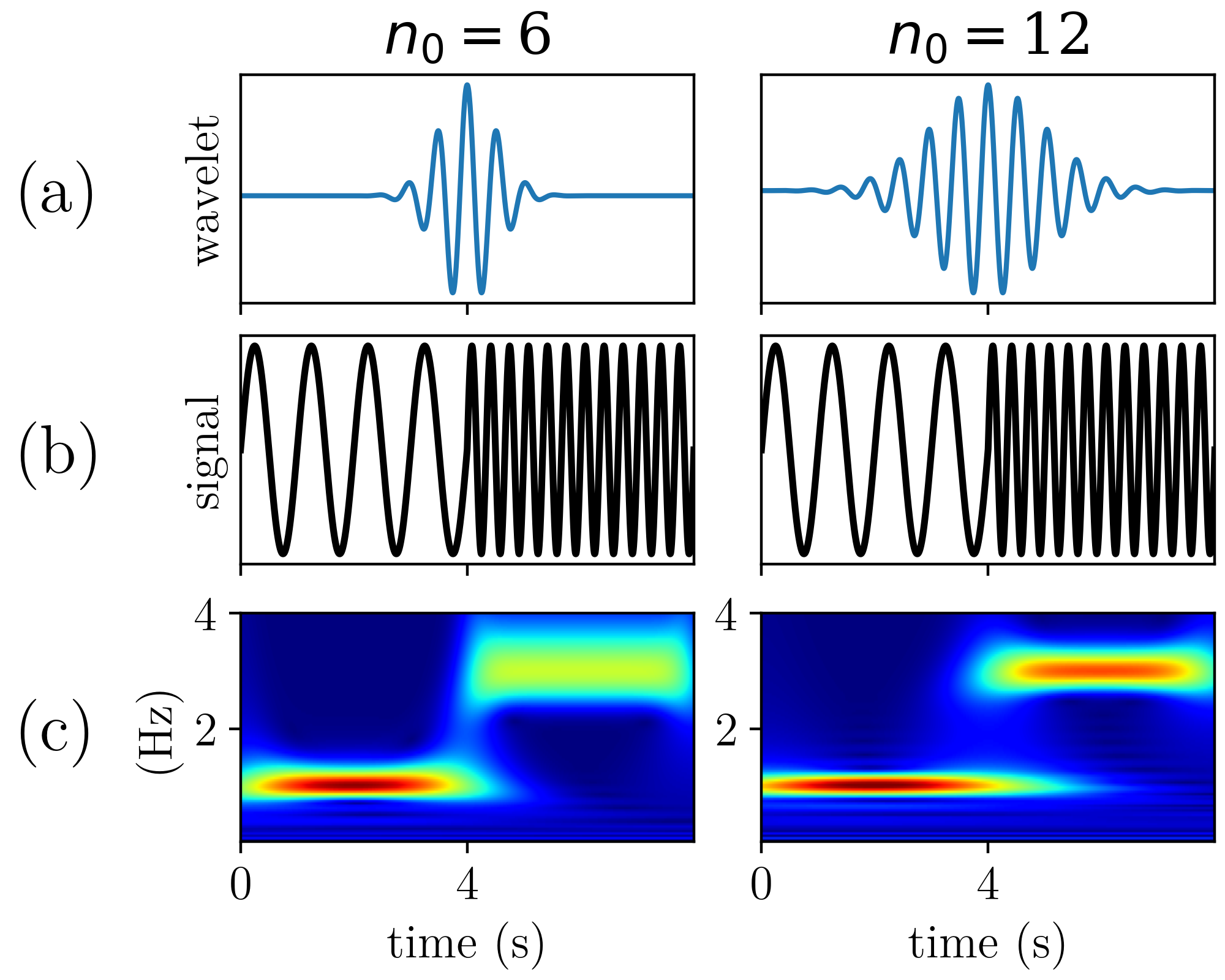}
    \caption{\textbf{Wavelet transform of a signal for different wave numbers: $n_0=6$ (left) and $n_0=12$ (right).}
    (a) Re\{$\Psi_{n_0}$\}.
    (b) Signal: \qty{1}{\hertz} sine function (\qtyrange[range-units=single,range-phrase=-]{0}{4}{\second}) combined to a \qty{3}{\hertz} sine function (\qtyrange[range-units=single,range-phrase=-]{4}{8}{\second}).
    (c) $|W(t,f)|$.}
    \label{fig:wavelet_n0}
\end{figure}

\begin{figure}[h!]
    \centering
    \includegraphics[width=0.4\linewidth]{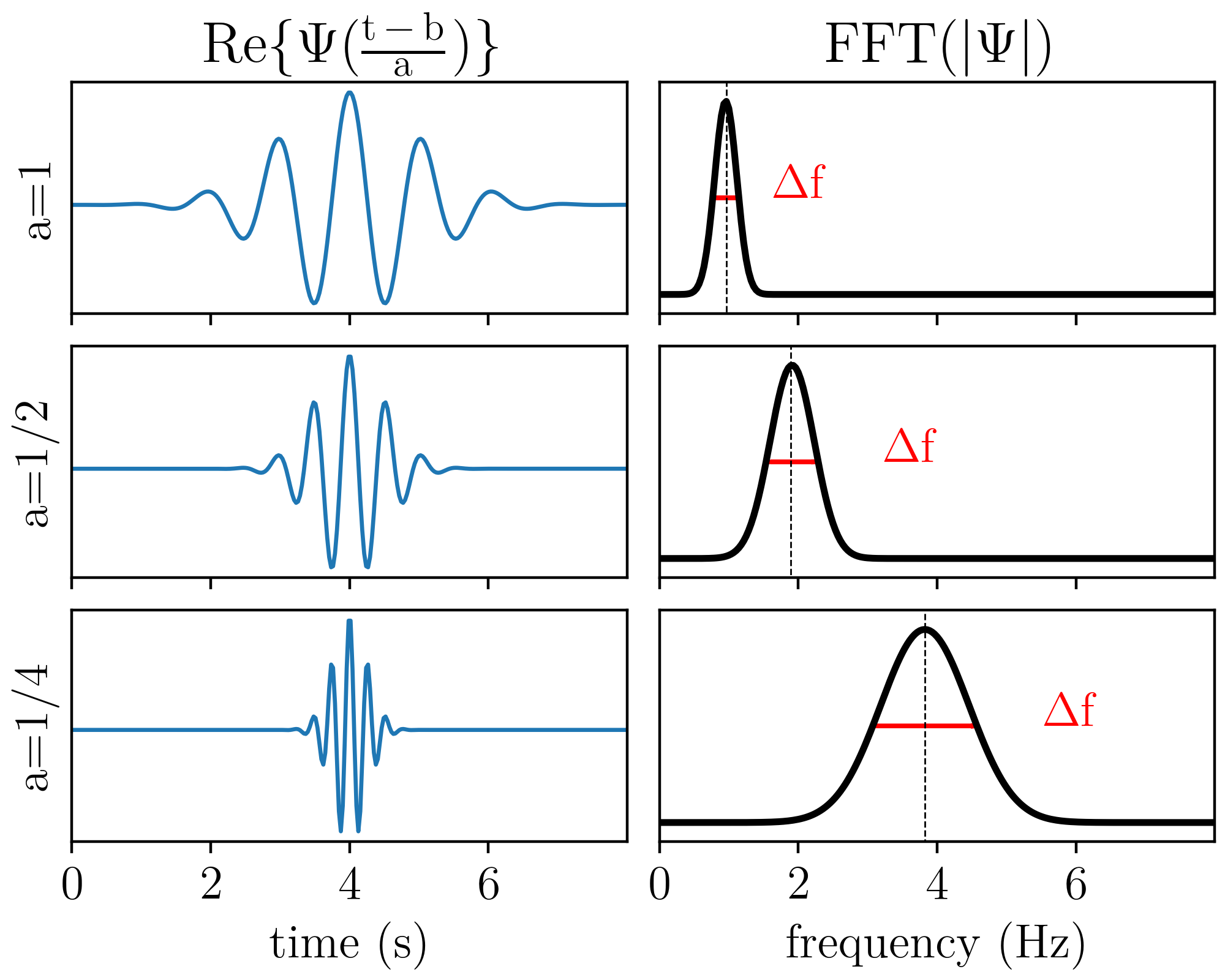}
    \caption{\textbf{Real part of the Morlet wavelet in time (left) and spectral (right) domain.} From top to bottom, the scale parameter varies from $a=f^*/f=1$ to $1/4$. The Morlet mother wavelet for CWT is a sine function convolved with a Gaussian function, given by $\Psi_{n_0} (t) = \pi ^ {-1/4} e^{in_0 t}e^{-t^2 / 2}$, where $n_0\geq 5$ represents the wave number of the wavelet, proportional to the number of oscillations of the mother wavelet in time \cite{torrence_practical_1998}. What distinguishes the wavelet transform from other time-frequency transforms is its constant quality factor $Q$ across frequencies (right panels). The $Q$-factor is defined by $Q=f^*/ \Delta f$ where $f^*$ is the frequency of the maximum of the mother wavelet in Fourier space and $\Delta f$ its Full-Width-Half-Maximum. Increasing $n_0$ results in a better frequency accuracy (higher $Q$-factor) at the expense of time accuracy.
    }
    \label{fig:morlet_qfactor}
\end{figure}

Here, we use the Continuous Wavelet Transform (CWT), which provides a time-frequency representation of a signal, {\em i.e.} the temporal changes of both its amplitude and its phase \cite{torrence_practical_1998}. 
For a real signal $S(t)$, the CWT yields the complex two-parameter function $W(b,a)$:
\begin{equation}
W(b,a) = \frac{1}{|a|^{1/2}} \int^{+\infty} _{-\infty} S(t) \overline{\psi}\left(\frac{t-b}{a}\right) {\rm d}t
\end{equation}
with $\psi(t)$ the mother wavelet, $a$ the scaling factor, and $b$ the translation factor.
$\overline{\psi}$ stands for the complex conjugate of $\psi$.
For convenience, in the remainder of the paper we use the equivalent notation $W(t,f)$, where the scale-frequency conversion is given by $f=f^*/a$, with $f^*$ being the frequency at which the Fourier transform of the mother wavelet, $\Psi(f)$, peaks. 
Therefore, $W(t,f)$ and $W(b,a)$ represent the same quantity, expressed respectively in the time-frequency and time-scale domains. The Continuous Wavelet Transforms were performed using the python library \textit{pycwt}. Since our simulated and experimental signals are close to sinusoidal, we use a Morlet mother wavelet, with a wave number $n_0$, made of a sine function convolved with a Gaussian function (Supp. Fig.~\ref{fig:wavelet_n0}) \cite{kronland-martinet_analysis_1987,torrence_practical_1998,cohen_better_2019}. What distinguishes the wavelet transform from other time-frequency transforms is its constant quality factor $Q$ across frequencies (Supp. Fig.~\ref{fig:morlet_qfactor}).
This makes wavelets especially suitable for analyzing oscillatory signals with time-varying frequency, such as the one from CR cells undergoing a photophobic response (Fig.~\ref{fig:exp_setup}(a,b)).

\subsection{Phase Synchronization Index: PSI}
To analyze the occurrence of synchronization, we extend the already published Phase Synchronization Index (PSI) \cite{schack_quantification_2005,liang_synchronization_2016} to its complex form.
From discrete phase signals $\{\varphi_1^1,\varphi_1^2,...,\varphi_1^N\}$ and $\{\varphi_2^1,\varphi_2^2,...,\varphi_2^N\}$, where $N$ refers to the size of the dataset, the PSI, noted $\widetilde{\Upsilon}$ reads:
\begin{equation}
    \widetilde{\Upsilon} = \left< e^{i(n\varphi_1^k - m\varphi_2^k )}\right>_k \; ,
    \label{eq:PSI_complex}
\end{equation}
where $k$ is a constant, and $n,m$ are positive integers that correspond to different harmonic factors.
$|\widetilde{\Upsilon}|$ ranges from 0 (no synchronization) and 1 (full synchronization).
In the following, we focus on the synchronization between fundamental modes: $n=m=1$, corresponding to a nearly resonance of the two oscillators. 

While this standard Phase Synchronization Index (PSI) analysis described above is powerful to study synchronization over time, it assumes that oscillations have stationary frequencies and is therefore limited when applied to transient or nonstationary signals.
In contrast, our refined approach described below extends the PSI to the time-frequency domain, providing precise insight into how synchronization evolves during rapid transitions such as photoshock.
We therefore introduce $\widetilde{\Upsilon}_{\Psi} (t,f)$, the PSI defined in the time-frequency domain by including the CWT of each signal, $W_1(t,f)$ and $W_2(t,f)$, in Eq. (\ref{eq:PSI_complex}), without the need of computing explicitly their phases, which yields:
\begin{equation}
\widetilde{\Upsilon}_{\Psi} (t,f) = \left< \frac{W_1(t,f) \overline{W_2}(t,f)}{|W_1(t,f)||W_2(t,f)|} \right>_{\Delta T _{\Psi}} = \Upsilon_\Psi \; e^{i\Phi} \; \label{eq:PSI_CWT_complex}.
\end{equation}
where the temporal averaging $\left<\cdot\right>_{\Delta T _{\Psi}}$ is performed over the wavelet window $\left[t-\frac{\Delta T _{\Psi}}{2},t+\frac{\Delta T _{\Psi}}{2}\right]$, with $\Delta T _{\Psi} \propto n_0 / f^*$ ensuring that the PSI is computed over a timescale appropriate to the analyzed frequency. 
In Supplementary Fig.~\ref{fig:SUP_PSI_on_sim}, we compare the PSI computation on a fixed size window (\qty{1}{\second}) to this adaptive size $\Delta T _{\Psi}$.
Following this framework, the phase and the modulus of $\widetilde{\Upsilon}_\Psi$ are variables of $t$ and $f$: $\Phi(t,f)$ and $\Upsilon_{\Psi} (t,f)$, respectively.

The time-frequency complex PSI can be compared to the time-frequency complex coherence that quantifies the linear coupling between two signals \cite{bruns_fourier-_2004,liang_synchronization_2016,guillet_tracking_2021}: 
\begin{equation}
    \kappa_{1,2} (t,f) = \frac{\left< W_1(t,f) \overline{W_2}(t,f)\right>_{\Delta T_{\Psi}}}{\sqrt{\left<|W_1(t,f)|^2\right>_{\Delta T_{\Psi}}\left<|W_2(t,f)|^2 \right>_{\Delta T_{\Psi}}}} \label{eq:Coherence_CWT} \;
\end{equation}
Perfect coherence between two signals at a given frequency occurs when they maintain both a constant phase difference and a constant amplitude ratio over the time interval considered. 
Eq.~(\ref{eq:Coherence_CWT}) differs markedly from Eq.~(\ref{eq:PSI_CWT_complex}). Specifically, in coherence, the averaging is performed separately on the numerator and denominator before computing the ratio. 
As a result, the coherence computation includes both amplitude and phase information from the full signal. 
In contrast, $\widetilde{\Upsilon}_\Psi$ is computed by normalizing out amplitude at each time-frequency point prior to averaging. 
This makes it particularly suited for detecting phase relationships independently of amplitude fluctuations. 
Importantly, if the signal amplitudes remain constant over time, as implicitly assumed in the derivation of PSI, then $\Upsilon_\Psi$ becomes equivalent to the modulus of the complex coherence $\kappa_{1,2}$. 

\section{Results\label{sec:Res}}

\subsection{Coupled oscillators simulations}

Fig.~\ref{fig:coupling_sim} illustrates the dynamics of a system composed of two coupled oscillators governed by Eq.~(\ref{eq:S_delta_noise}), with fixed natural frequencies $\omega_1/2\pi =\qty{30}{\hertz}$, $\omega_2/2\pi = \qty{20}{\hertz}$. The coupling strength $B$ and initial phases $\varphi_1(0)$, $\varphi_2(0)$ vary between cases.
Steady-state synchronization is reached when the phase difference becomes constant over time, i.e. $\dot{\delta}=\dot{\varphi _1} - \dot{\varphi _2}=0$. 
Analytically, based on Adler's equation given above, this condition is met in the limit $t\rightarrow \infty$, which means that a sufficiently strong coupling exists $|2B|>|\Delta\omega|$. 
In such a strong coupling regime (Fig.~\ref{fig:coupling_sim}(a)), the two oscillators quickly synchronize (left panel) with a constant phase difference (right panel) indicating in-phase synchronization ($\delta=0$). 
In contrast, in a weak coupling condition the system does not maintain synchronization ($|2B|<|\Delta\omega|$, Fig.~\ref{fig:coupling_sim}(b)). 
Although transient locking occurs, the oscillators remain predominantly desynchronized, with a quasi periodic behavior. 
Note that the sign of $B$ determines the nature of the synchronization: the two oscillators can synchronize in phase if $B$ is negative (Fig.~\ref{fig:coupling_sim}(a)) or in anti-phase if $B$ is positive ($\delta=\pi$, Fig.~\ref{fig:coupling_sim}(c)). 
Furthermore, even in the synchronized time-region, the instantaneous frequencies of the oscillators vary. This occurs because the synchronization frequency $\omega_S = \dot{\varphi _1} = \dot{\varphi _2} = {(\omega_1 +\omega_1)}/{2}$ generally differs from the natural frequency of each uncoupled oscillator. 
These transient dynamics highlight the importance of time-frequency methods to detect synchronization in nonstationary systems.

\begin{figure*}[!t]
    \centering
    \includegraphics[width=1\linewidth]{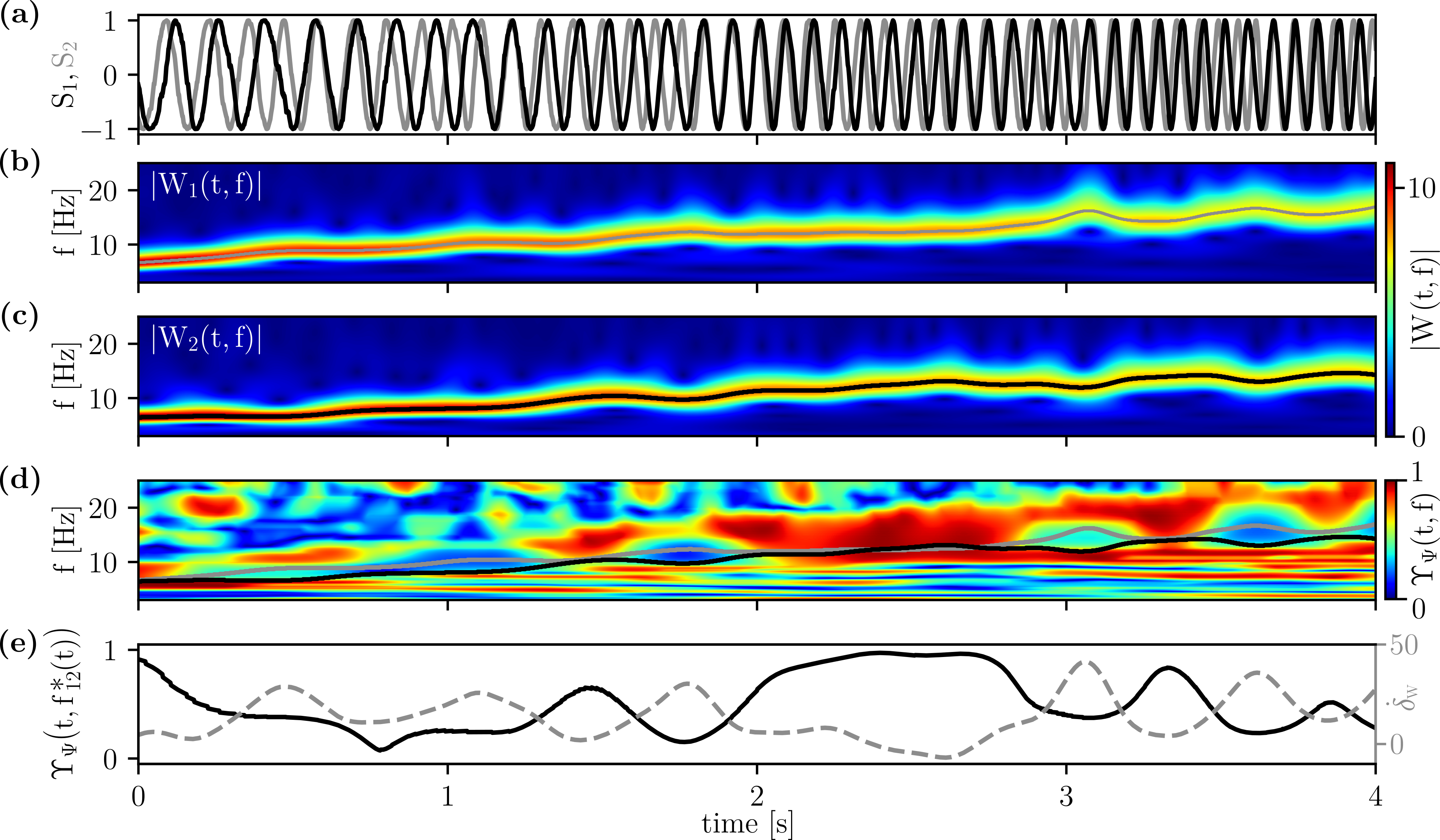}
    \caption{\textbf{System of two coupled oscillators with linearly increasing frequency.}
    (a) Simulated signals generated from Eq.~(\ref{eq:S_delta_noise}). Gray: $\cos(\varphi_1)$; black: $\cos(\varphi_2)$. Frequencies: $\omega_1/2\pi=5Hz$, $\omega_2/2\pi=4Hz$.
    Coupling strength: $B_1/2\pi=B_2/2\pi=0.5Hz$
    (b) $|W_1(t,f)|$ for the first oscillator (gray line in (a)) using Morlet wavelet with $n_0=6$. The gray line indicates the local frequency maxima $f_1^*(t)$. 
    (c) $|W_2(t,f)|$ for the second oscillator (black line in (a)) using Morlet wavelet with $n_0=6$. The black line indicates the local frequency maxima $f_2^*(t)$. 
    (d) $\Upsilon_\Psi(t,f)$ computed with an adaptive time window $\Delta T_\Psi \propto n_0/f^*$ ($n_0=6$). Frequency ridges $f_1^*(t)$ (gray) and $f_2^*(t)$ (black) are overlaid.
    (e) $\Upsilon_\Psi \left(t,f^*_{12}(t)\right)$ (black), computed along the frequency trajectory $f^*_{12}(t) = (f_1^*(t) + f_2^*(t))/2$. The derivative of the phase difference between $S_1$ and $S_2$, $\dot\delta_W (t)$ (gray) was computed using a Gaussian derivative wavelet  to filter out the signal noise (Supplementary Fig.~\ref{fig:wavelet_derivative}) \cite{mallat_wavelet_2008}. }
    \label{fig:PSI_on_sim}
\end{figure*}

To better illustrate nonstationary dynamics, in addition of a positive coupling, we now define a system of two chirp oscillators whose natural frequencies increase linearly over time (Fig.~\ref{fig:PSI_on_sim}(a)). 
This supposedly mimics realistic biological or physical systems where oscillator properties evolve gradually due to internal modulation or external perturbations. For example, circadian rhythms adjust their phases and periods in response to light-dark cycles \cite{pittendrigh_functional_1976}, or cardiac pacemakers can shift frequency via modulatory feedback \cite{brown_cardiac_1979}.

We pick an intermediate value for coupling ($B=2\Delta\omega$) in Eq.~(\ref{eq:S_delta_noise}) to favor longer synchronization states.
The spectrograms of the two oscillators are shown in Fig.~\ref{fig:PSI_on_sim}(b-c), where black and gray lines display the instantaneous frequency $f^*(t)$, obtained as the local maxima of the wavelet modulus, also called  ``ridges'', which verify $\partial_f|W(t,f^*)|=0$ \cite{carmona_identification_1995}.

Obviously, this computation enables us to follow the linear increase of each oscillator frequency. 
While these two signals are mostly synchronized, few phase slips are visible on the time-domain signals (Fig.~\ref{fig:PSI_on_sim}(a)) (from \qty{3}{\second} to \qty{3.2}{\second} for instance). 
The time-frequency maps confirm the presence of transient phase slips, displaying frequency mismatch (lines in Fig.~\ref{fig:PSI_on_sim}(d)). 

Fig.~\ref{fig:PSI_on_sim}(d) shows the time–frequency map of the PSI $\Upsilon_\Psi(t,f)$ for the two coupled chirp oscillators.
For reference, the instantaneous frequency ridges $f_1^*(t)$ (gray) and $f_2^*(t)$ (black), previously extracted in Fig.~\ref{fig:PSI_on_sim}(b–c), are overlaid as guide lines.
Regions of low $\Upsilon_\Psi$ appear as localized ``islands'' in this map, marking transient losses of synchronization that are clearly bounded by the main frequency ridges.

\begin{figure}[h!]
    \centering
    \includegraphics[width=.9\linewidth]{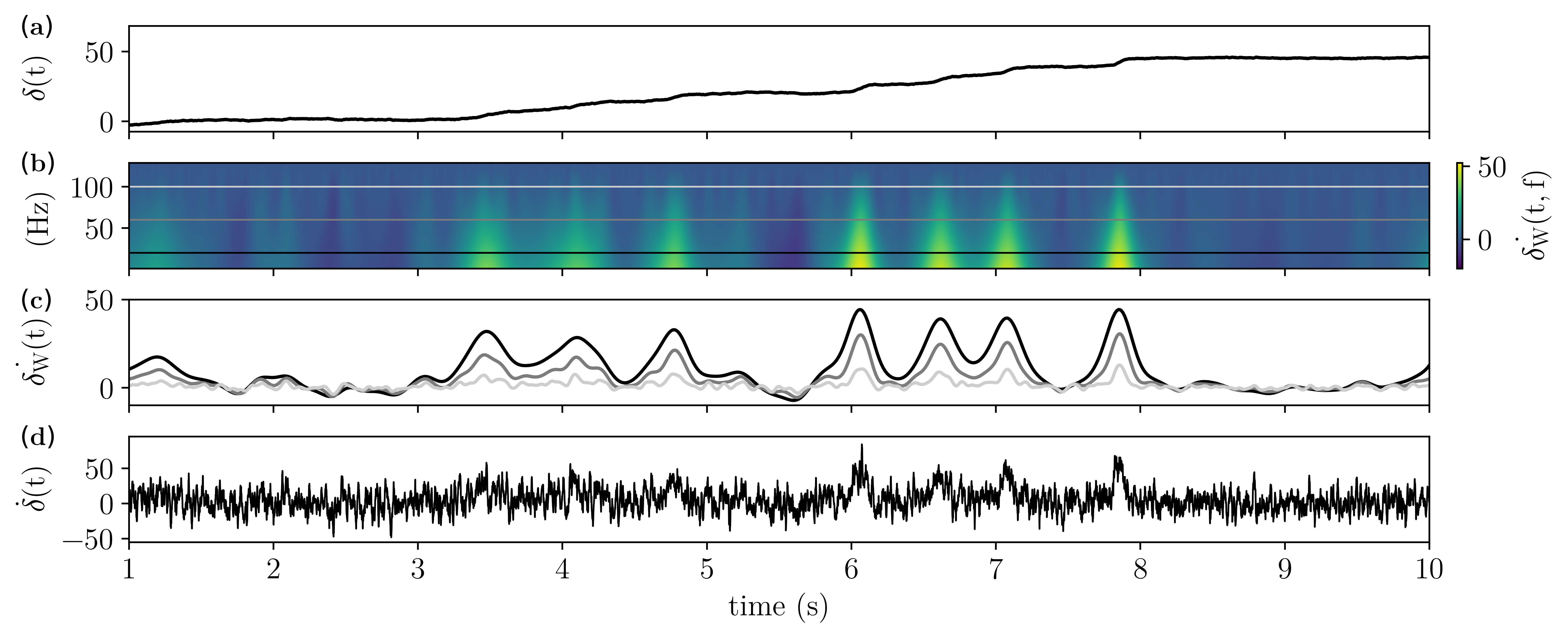}
    \caption{\textbf{Temporal derivative of the $\delta$ computed using a Gaussian derivative wavelet \cite{mallat_wavelet_2008}.}
    (a) $\delta(t)$ using Fig.~\ref{fig:PSI_on_sim} signal.
    (b) Wavelet transform modulus  of $\delta (t)$ shown in (a) with a Gaussian derivative $\psi_{\dot{G}} $ mother wavelet. We have the relation:  
    \newline
    \hspace*{7.cm}$W_{\psi_{\dot{G}}} [\delta](t,f) \propto  W_{\psi_G} [\dot{\delta}](t,f) $.
     \newline
    The horizontal lines correspond to different frequencies (20, 60, and 100 Hz).
    (c) Profiles $\dot{\delta}_W(t) = W_{\psi_{\dot{G}}} [\delta](t,f)$ for the frequencies displayed in (b).
    (d) Classical derivative of the signal $\dot{\delta}(t) \sim {(\delta(t+{\rm d}t)-\delta(t))}/{{\rm d}t}$ with ${\rm d}t=\qty{10}{\milli\second}$. We note that the scale of the wavelet changes the amplitude of the phase jumps (singularities of $\delta$), the smaller the scale ($a f^*/f$), the sharper the local extrema of $\dot{\delta}$. However the temporal localization of these jumps can still  be achieved for larger wavelet scales. 
    }
    \label{fig:wavelet_derivative}
\end{figure}

To quantify the synchronization dynamics more directly, we compute $\Upsilon_\Psi\left(t,f^*_{12}(t)\right)$ along the averaged frequency trajectory $f^*_{12}(t) = \left[f_1^*(t)+f_2^*(t)\right]/2$ (black line, Fig.~\ref{fig:PSI_on_sim}(e)).
On the same plot, we display the temporal derivative of the phase difference between the two oscillators, $\dot{\delta}_W(t)$ (gray curve), which provides a complementary measure of phase locking, and therefore a validation of phase synchronization.
Both quantities were computed directly in the wavelet domain: $\dot{\delta}_W(t)$ was obtained using a Gaussian derivative wavelet, which efficiently filters out high-frequency noise while preserving the temporal localization of phase jumps (Supplementary Fig.~\ref{fig:wavelet_derivative}).

\begin{figure}[t!]
    \centering
    \includegraphics[width=.9\linewidth]{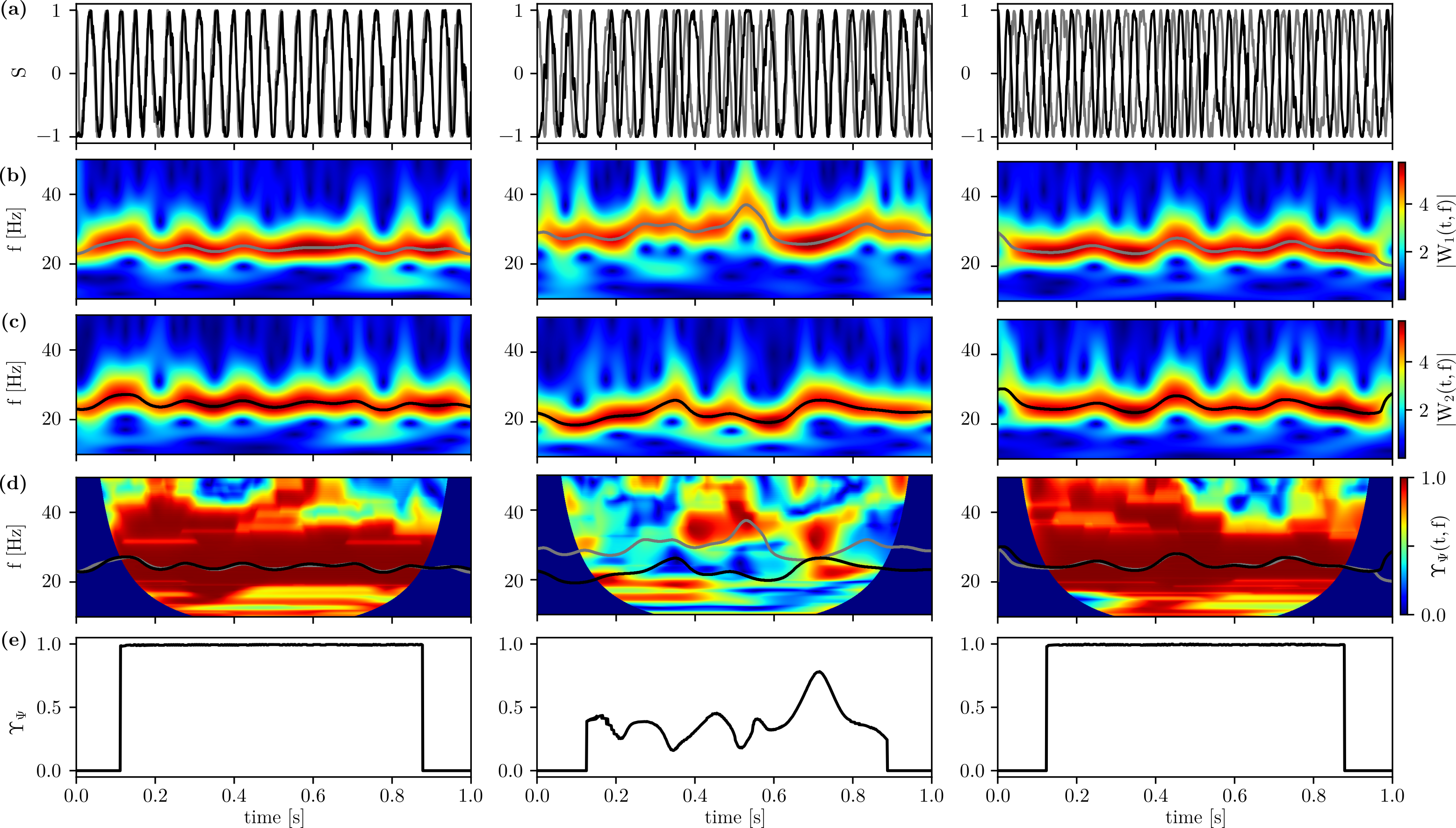}
    \caption{\textbf{System of two coupled oscillators with  Gaussian noise, modeled by Eq.~(\ref{eq:S_delta_noise}).}
    Left column corresponds to strong negative coupling ($B/2\pi=\qty{-20}{\hertz}<0$, $|2B|>\Delta \omega$), with $\varphi_1 (0) = 0$, $\varphi_2 (0) = \pi$: leads to stable in-phase synchronization (same data as Fig.~\ref{fig:coupling_sim}(a)). 
    Central column corresponds to weak negative coupling ($B/2\pi=\qty{-3}{\hertz}<0$, $|2B|<\Delta \omega$), with same initial phases: leads to desynchronization with transient coordination (same data as Fig.~\ref{fig:coupling_sim}(b)).
    Right column corresponds to strong positive coupling ($B/2\pi=\qty{20}{\hertz}>0$, $|2B|>\Delta \omega$) with $\varphi_1 (0) =\varphi_2 (0) = 0$: leads to stable anti-phase synchronization. (same data as Fig.~\ref{fig:coupling_sim}(c)).   
    (a) Simulated signal following Eq.~(\ref{eq:S_delta_noise}). Gray: $\cos(\varphi_1)$. Black: $\cos(\varphi_2)$.
    (b) $|W_1(t,f)|$, $n_0=6$. Gray line: local frequency maxima $f_1^*(t)$.
    (c) $|W_2(t,f)|$, $n_0=6$. Black line:  local frequency maxima $f_2^*(t)$.
    (d) $\Upsilon_\Psi(t,f)$ computed with a variable window $\Delta T_\Psi = n_0/f^*$ (Eq.~\ref{eq:PSI_CWT_complex}), $n_0=6$. Gray: $f_1^*(t)$. Black: $f_2^*(t)$.
    (e) Plain line: $\Upsilon_\Psi (t,f^*_{12}(t))$ where $f^*_{12}(t) = (f_1^*(t) + f_2^*(t))/2$. Note that the influence cone is visible in panels (d-e), more details in \cite{torresani_analyse_1995}.}
    \label{fig:coupled_sim_noisy_PSI}
\end{figure}

The comparison between $\Upsilon_\Psi$ and $\dot{\delta}_W$ in Fig.~\ref{fig:PSI_on_sim}(e) confirms that regions of strongest synchronization ($\Upsilon_\Psi \to 1$) correspond to the lowest phase-derivative values ($\dot{\delta}_W \to 0$).
For comparison, Supplementary Fig.~\ref{fig:SUP_PSI_on_sim} shows $\Upsilon_\Psi$ computed with a constant time window; in this case, the ``islands'' of desynchronization vanish, and synchronization states are poorly resolved, demonstrating the advantage of the adaptive formulation.
Furthermore, applying this framework to the coupled oscillator model of Fig.~\ref{fig:coupling_sim} reproduces the same qualitative behavior (Supplementary Fig.~\ref{fig:coupled_sim_noisy_PSI}), confirming the robustness of the method.
Together, these results show that the proposed time-frequency PSI method provides an accurate and temporally resolved characterization of synchronization dynamics, thereby motivating its application to biological oscillators.

\subsection{Flagellar resynchronization in response to a photophobic stimulus}

\begin{figure*}[!t]
    \centering
    \includegraphics[width=1\linewidth]{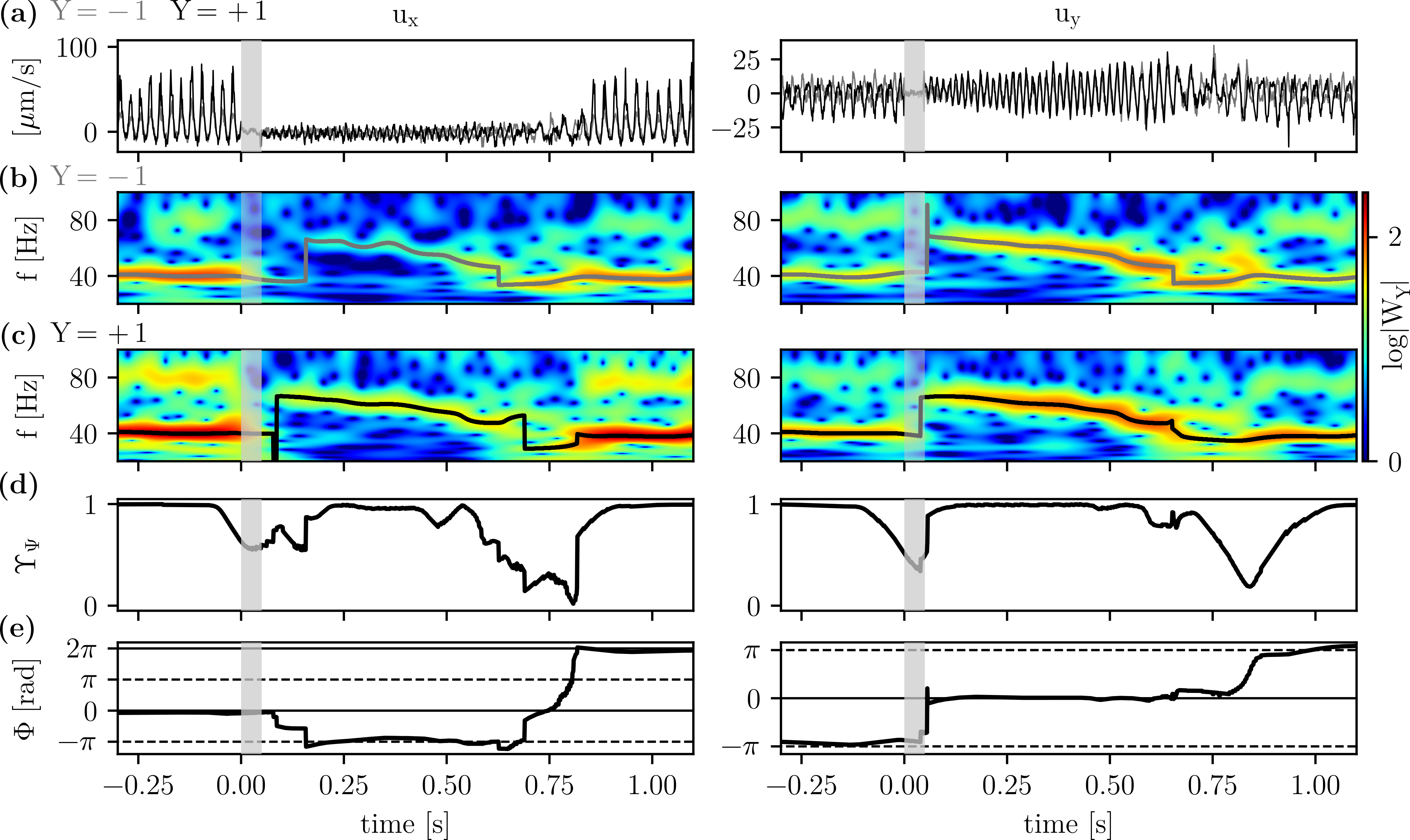}
    \caption{\textbf{Flagellar resynchronization of CR following a photoshock.}
    (a) Averaged fluid velocity signals along the $x$-axis (left) and $y$-axis (right), extracted from regions on each side of the alga: $Y=-1$ (gray) and $Y=+1$ (black), as shown in Fig.~\ref{fig:exp_setup}.
    (b-c) Time-frequency representation of the CWT signals, $\log |W_Y(t,f)|$ (Morlet wavelet with $n_0=12$), for (b) $Y=-1$ and (c) $Y=+1$. Frequency ridges $f_{-1}^*(t)$ (gray) and $f_{+1}^*(t)$ (black) are also plotted.
    (d) Phase Synchronization Index, $|\widetilde{\Upsilon}_\Psi| (t, f^*_{\pm 1}(t))$, computed at the mean instantaneous frequency $f^*_{\pm1}(t) = [f_{-1}^*(t)+f_{+1}^*(t)]/2$.
    (e) Phase $\Phi(t)$ obtained from the complex PSI $\widetilde{\Upsilon}_\Psi$.
In all panels, the shaded gray region marks the duration of the photoshock.}
    \label{fig:resync}
\end{figure*}

The time response of CR to a photoshock unfolds in three distinct phases (Fig.~\ref{fig:resync}(a)). 
First, before stimulation ($t<\qty{0}{\milli\second}$), the cell swims in a breaststroke motion in which the two flagella are synchronized, resembling two noisy oscillators with strong coupling, as described in Fig.~\ref{fig:coupling_sim}(a,c). The velocities recorded along the $x$-axis are in phase, while the ones along along $y$-axis are in anti-phase.
Then, immediately after the photoshock ($\qty{0.05}{\second}<t\lesssim\qty{0.6}{\second}$), the beating switches to an undulatory mode, which drives backward swimming with smaller amplitude. In this regime, the two flagella oscillate in phase along the $y$-axis.
It should be noted that during this phase, the signals detected at $Y=\pm1$ reflect the combined flow generated by both flagella, which can no longer be distinguished individually. 
However, once a flagellum resumes its breaststroke-like motion, the signal from the nearest window recaptures its specific dynamics, while the opposite window no longer detects it.
Lastly, for $t\gtrsim\qty{0.6}{\second}$, the breaststroke pattern is progressively restored, until full resynchronization.

To quantitatively characterize the dynamics of the transitions between these different stages, we apply our wavelet-based time-frequency analysis to the velocity signals (Fig.~\ref{fig:resync}(b-c)).
Specifically, for each lateral observation window located at $Y\pm1$, we compute the continuous wavelet transforms, denoted $W_{\pm1}(t,f)$.
This approach allows us to clearly identify the three behavioral phases described above, while providing detailed insight into the instantaneous frequency evolution of the flagellar beating.
Both $x$- and $y$-axis velocity components (left and right panels, respectively) show qualitative similar trends, while their relative amplitudes reflect the dominance of different swimming modes at each stage. Note that the measured velocity signals do not directly reflect flagellar kinematics, but rather the surrounding flow field generated by flagellar beating. The observed spectral content thus results from the fluid velocity generated by the time-dependent flagellar waveform, rather than from a direct decomposition of flagellar deformation modes.

The wavelet modulus maxima frequency (marked with lines on the spectrograms of Fig.~\ref{fig:resync}(b-c)) follows a reproducible sequence. 
In the pre-stimulus phase, both flagella exhibit steady beating frequencies centered around \qty{40}{\hertz} (Fig.~\ref{fig:resync}(b-c)).
Following the photoshock, the beating frequency rises sharply to \qty{70}{\hertz}, consistent with the undulatory backward-swimming mode.
Because this mode primarily generates flow along the $y$-axis, its detection is more robust in the $y$-velocity component. In contrast, along the $x$-axis the associated energy is weaker, which can result in a delayed or less sharply defined ridge during the transition. In this case, the \qty{70}{\hertz} band is still present in the spectrogram but carries lower energy, and the local maximum of the wavelet modulus may be more sensitive to reduced signal-to-noise ratio (left panels in Fig.~\ref{fig:resync}b).
Similarly, transient post-photoshock undershoots ((Fig.~\ref{fig:resync}(c, left panel)), and overshoots (Fig.~\ref{fig:resync}(b, right panel)), originates from ridge-tracking uncertainty when the wavelet modulus is locally low. These features reflect numerical artifacts of the ridge extraction in low-energy regions.

Immediately following the photoshock, the dominant mode gradually decreases to around \qty{50}{\hertz} over several hundred milliseconds.
This transition is followed by a sharp drop into a low-frequency regime, approximately \qty{30}{\hertz}, marking the start of resynchronization.
The beating frequency then progressively increases and stabilizes near the pre-stimulus breaststroke value, indicating the recovery of the original synchronized state.

\begin{figure}[!t]
    \centering
    \includegraphics[width=1\linewidth]{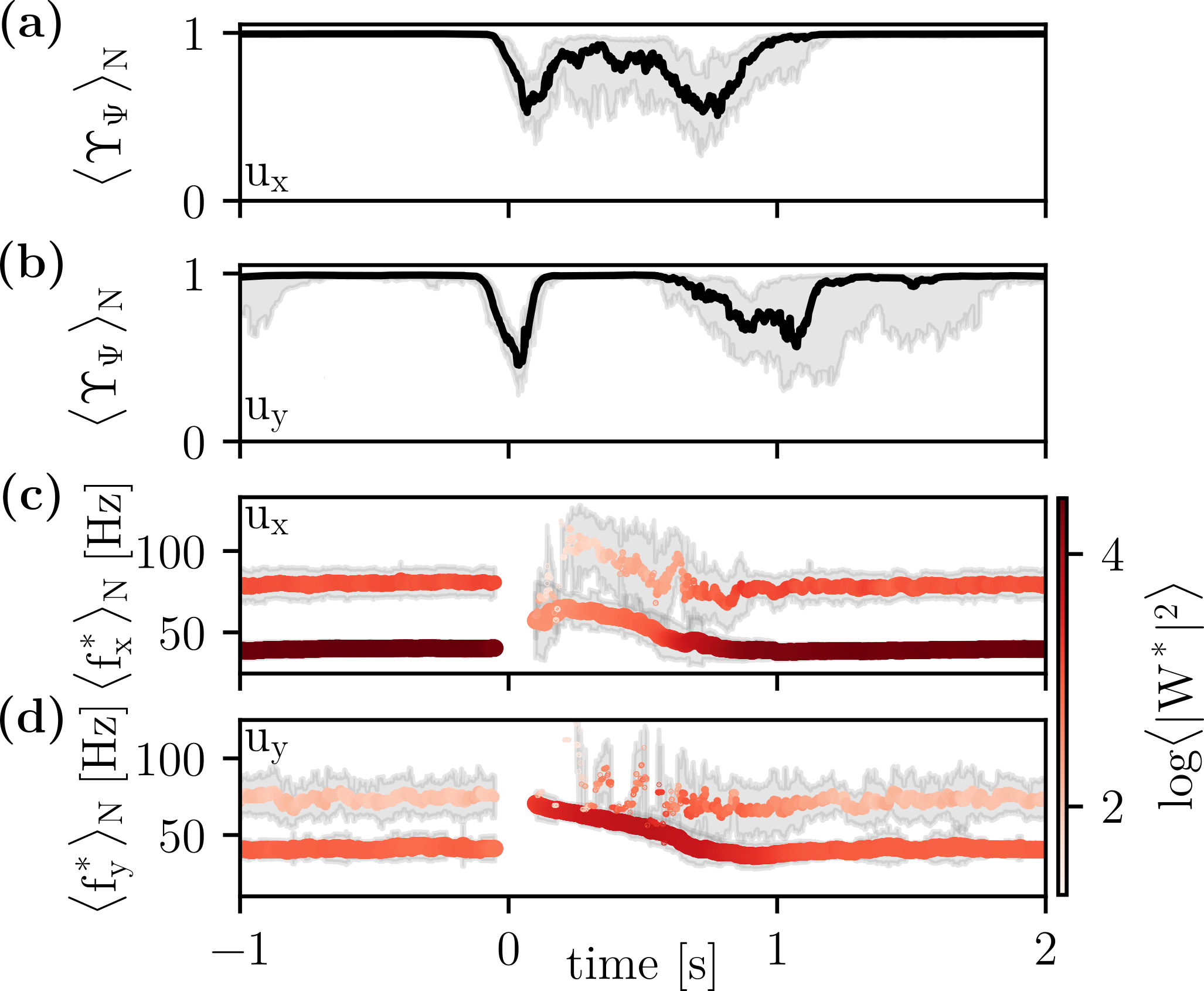}
    \caption{\textbf{Flagellar synchronization and frequency dynamics around photoshock.}
    (a-b) PSI averaged over 44 photoshocks (13 algae) computed using the two signals $Y=-1$ and $Y=1$, along the $x$-axis (a) and the $y$-axis (b). Black line: Median PSI. Gray shade: First quartile (Q1) and Third quartile (Q3).
    (c-d) The red lines correspond to the averaged wavelet frequency maxima of each band across all the algae, computed using signal in $Y=0$. The gray shades are the standard deviations. The color map is linked to the amplitude and represents the average wavelet transform squared modulus of the corresponding frequency maxima, in log scale.
    Frequencies between $t=-\qty{50}{\milli\second}$ and $t=\qty{100}{\milli\second}$ are not shown because they are not robustly detected using the wavelet transform.
    }
    \label{fig:freq_maxima}
\end{figure}

To quantify phase synchronization more precisely, we apply our PSI framework to these flagellar signals, using Eq.~(\ref{eq:PSI_CWT_complex}). 
Instantaneous frequency maxima $f^*_{-1}$ and $f^*_{+1}$ were extracted from $|W_{-1}(t,f)|$ and $|W_{+1}(t,f)|$, respectively (lines in Fig.~\ref{fig:resync}(b-c)). 
The subscripts $\pm1$ refer to the averaging windows located on the two opposite sides of the cell body ($Y=\pm1$; see Fig.~\ref{fig:exp_setup}(a)).
$\Upsilon_\Psi \left(t,f^*_{\pm1}(t)\right)$ is evaluated using an adaptive time window such as $f^*_{\pm1}(t)= (f^*_{-1}+f^*_{+1}) / 2$ (Fig.~\ref{fig:resync}(d)).
This analysis displays two desynchronization events: immediately following the photoshock (from \num{0} to \qty{.2}{\second}), and after the stimulus, roughly from \num{.5} to \qty{1}{\second}.
While the first is linked to an abrupt change in flagellar beating modes, the second reflects a subtler change in phase synchronization.
This is further confirmed by the PSI phase $\Phi(t,f^*_{\pm 1})$, computed as the phase of the complex $\widetilde{\Upsilon}_\Psi$ (Equation ~(\ref{eq:PSI_CWT_complex})), which transitions along $y$ from in-phase ($\Phi \approx 0$) to anti-phase ($\Phi \approx \pi$) during the recovery (Fig.~\ref{fig:resync}(e)).
This gradual shift highlights reorganization of the underlying coupling dynamics between the two flagella.

Finally, while Fig.~\ref{fig:resync} shows a representative case, we confirmed robustness across the dataset by averaging $\Upsilon_\Psi(t,f^*_{\pm1})$ over all recorded photoshock events (Fig.~\ref{fig:freq_maxima}(a,b)), retaining only experiments with strong pre- and post-shock synchronization. A photoshock response is thus included in the data set only if the time-averaged synchronization index satisfies $\left<\Upsilon_\Psi\right> > 0.85$, where the averaging is performed outside the photophobic response window. Specifically, denoting by $t_{\rm pc}$ the start of the light pulse, we exclude the interval $\left[t_{\rm pc}-0.1\unit{\second},t_{\rm pc}+1,5\unit{\second}\right]$, and compute the average over the remaining time points. Photoshock events that did not meet this criterion typically corresponded to cases where synchronization was already weak before stimulation (e.g. noisy beating or partial desynchronization). The lack of clear pre-shock synchronization makes it difficult to meaningfully average synchronization across such events.

Beyond tracking fundamental frequencies, the time-frequency maps in Fig.~\ref{fig:resync}(b-c) also reveal higher-harmonic structure in the flagellar beating.
Before the shock, several bands are visible on the spectrograms, the two dominant ones corresponding to a fundamental at $\sim$\qty{40}{\hertz} and its first harmonic at $\sim$\qty{80}{\hertz} (Fig.~\ref{fig:resync}(b,c)). To systematically characterize the dynamics of these harmonics across photoshock events, we quantified the evolution of the two dominant modes in terms of both frequency and amplitude.
Specifically, instantaneous frequency ridges were extracted from the wavelet spectrograms, and only local maxima of the wavelet modulus exceeding a fixed threshold ($|W_Y|>5$) were retained (Fig.~\ref{fig:freq_maxima}(c–d)). This phenomenological criterion suppresses spurious ridge detections in low-energy regions and ensures that only energetically significant modes are analyzed.
While other frequency bands are also present in Fig.~\ref{fig:resync}(b-c), Fig.~\ref{fig:freq_maxima}(c-d) only report the two dominant modes with the highest energy.

Immediately following the photoshock, the spectral balance between these modes changes markedly. Along the $y$-axis, the mode that carries the most of the energy coincides with the first harmonic observed during the stationary breastroke regime (Fig.~\ref{fig:freq_maxima}d). 
Similarly, despite a lower signal-to-noise ratio along the $x$-axis, Fig.~\ref{fig:freq_maxima}c shows that immediately after the photoshock, the dominant frequency is shifted upward relative to the stationary regime, approaching the higher-frequency mode observed prior to the shock. 
Overall, along both directions, the relative amplitude and frequencies of these modes gradually relax toward their pre-stimulus values, over several hundred milliseconds until breaststroke swimming is restored.

These observations suggest a subtle mechanism in which higher harmonic components, already present during breaststroke beating, become dominant during photoshock. Rather than switching between discrete beating patterns, the cell appears to modulate the relative weight of coexisting oscillatory modes. Our results indicate that flagellar beating involves multiple coexisting spectral components whose relative amplitudes and frequencies change upon the photoshock, consistent with a redistribution of energy between coexisting swimming modes.

\section{Discussion\label{sec:Disc}}

Using a time-frequency framework combined with a wavelet-based Phase Synchronization Index (PSI), we investigated the photophobic response of \textit{Chlamydomonas reinhardtii} cells held in place by a micropipette (Fig.~\ref{fig:exp_setup}).
Our primarly goal was to establish and validate a time-frequency synchronization framework on a well-controlled biological response rather than to perform a systematic exploration of photoshock parameter space. For this reason, we focused on a single stimulus protocol (50 ms, fixed intensity) that reliably triggers photoshock. Exploring such parameter dependence would be a valuable extension of the present study, to quantitatively investigate whether the response is graded and whether long-term adaptation to light could be examined.

Our approach enabled us to follow the evolution of flagellar beating across different swimming stages, with high temporal and spectral resolution, and to capture how synchronization breaks and re-emerges. 
Our analysis shows that flagellar coordination does not switch abruptly between discrete modes, but instead involves a gradual redistribution of energy across coexisting oscillatory components.
In particular, we find that the fundamental breaststroke frequency temporarily becomes negligible during photoshock, while the higher-frequency mode becomes dominant, before both modes re-establish during resynchronization.
This could be interpreted by the fact that over time the algae modulate the strength of at least two swimming mechanisms before, during and after the strong light perturbation.
The persistence of these harmonic modes, even during breaststroke beating, may provide a ``spectral reserve'' that facilitates rapid switching into backward swimming under stress.

So far, our PSI analysis has focused on the fundamental modes ($n=m=1$; Eq.~\ref{eq:PSI_complex}).
This restriction becomes limiting during mode transitions. As illustrated in Fig.~\ref{fig:resync}(b–c), during the transition following the photoshock, the two oscillators do not switch modes simultaneously. In particular, along the $x$-direction, the ridge associated with $Y=-1$ transitions to the higher-frequency mode later than that of $Y=+1$. During this interval, the instantaneous frequencies satisfy $f_1^* / f_2^* \neq 1$. Computing the PSI under the assumption $n = m = 1$ therefore violates the near-resonance condition underlying the phase dynamics. As a consequence, the inferred phase difference might present artificial plateaus (e.g., near $-\pi/2$). These artifacts would be avoided by using one oscillator as a fixed reference rather than combining both instantaneous phases.
The generalized formulation detailed in the Supplemental Materials (Eq.~\ref{eq:PSI_CWT_complex_generalized}) resolves this issue by rescaling the phases according to the instantaneous frequency ratio, i.e., by choosing appropriate integers $(n,m)$ that satisfy $f_1^*/f_2^* \approx m/n$. Applying this correction suppresses the spurious phase jump observed immediately after the photoshock, as shown in Fig.~\ref{fig:new_PSI_generalized}.
The generalized method could be readily extended to probe synchronization between higher-order components ($n,m>1$), and test whether harmonics themselves are phase-locked, and how their relative weights evolve during photoshock.
This richer spectral organization resonates with studies suggesting that multiple internal oscillatory states coexist in CR flagella beating \cite{wan_coordinated_2016,friedrich_hydrodynamic_2016}.

At a broader level, our measurements are based on flow fields generated by the two flagella rather than their direct motion.
As a result, the recorded signals represent an effective measure of flagellar coordination but do not allow us to isolate the detailed contributions of each flagellum or identify the specific coupling pathways responsible for resynchronization.
Direct imaging of flagellar motion, ideally resolving \textit{cis} and \textit{trans} beating patterns, would help bridge this gap and allow a more detailed comparison with models that explicitly include amplitude dynamics and hydrodynamic interactions \cite{brumley_flagellar_2014,liu_transitions_2018,friedrich_hydrodynamic_2016,zorrilla_role_2025,hu_multiflagellate_2024}.
Importantly, the same wavelet-based extraction of instantaneous frequency and phase, and the corresponding Phase Synchronization Index, could be applied without modification to any direct flagellar signal, including curvature, tip displacement or tangent angle time series. Direct flagellar tracking would be a valuable next step.
A promising future application of our framework would be to apply our framework to the \textit{ptx1} mutant, which exhibits spontaneous transitions between in-phase and anti-phase modes \cite{zorrilla_role_2025}. Such dynamics would allow probing intrinsic synchronization and phase-slip events without external perturbations.
Interestingly, this would perhaps lead us to quantify the energy loss or recovery of algae in the different swimming stages: does the algae select a swimming mode on the basis of a criterion of economy or a criterion of adaptation? 
Such analyses would enable direct investigation of how coupling strength and frequency asymmetry influence synchronization and gait selection.

Together, these results demonstrate the power of adaptive time-frequency methods to resolve transient synchronization dynamics. 
They also highlight new questions on the interplay between intrinsic oscillatory modes, amplitude regulation, and coupling mechanisms in eukaryotic flagella.
While demonstrated here on flagellar dynamics, this approach may broadly be applied to other biological or physical systems where oscillatory coordination evolves in time.

\section{Acknowledgments}
We thank Ahmad Badr for insightful discussions. 
The authors acknowledge support of the Leverhulme Trust (grant RPG-2018-345; A.A. and M.P.) ; the SMR Department and CNRS Tremplin@INP (A.A.) ; MICINN (M.P.: grant PID2019-104232BI00/AEI; J.A.: grant PID2022-143018NB-I00/AEI).  The authors also acknowledge financial support from the Agence Nationale de la Recherche under Babin (ANR-25-CE30-3613-01).
M.P. acknowledges the fact that IMEDEA is an accredited `Mar\'ia de Maeztu Excellence Unit' (grant CEX2021-001198, funded by MCIN/AEI/ 10.13039/ 501100011033). 

A.A., J.A. and M.P. conceived the project. A.A. and J.A. designed and performed the experiments. L.F., A.A. and J.A. developed the analysis code. L.F. and F.A. designed and performed the signal analysis work. L.F., A.A. and F.A. wrote the manuscript with feedback from all authors.

The data and code supporting this work are available in Zenodo (\url{https://doi.org/10.5281/zenodo.17406887}).

\bibliographystyle{ieeetr}
\bibliography{ref_AA}

\pagebreak
\onecolumngrid 
\newpage

\supplementarysection
\section{Supplementary Materials\label{sec:SuppMat}}

\subsection{Flow field}

\subsection{Python packages used for the simulations}

The Python modules used for simulation and analysis are \textit{numpy, scipy, scipy.integrate-odeint, sdeint, pycwt, PyWavelets, matplotlib, pandas.}

\subsection{Morlet Wavelet transform}

\subsection{Time frequency PSI in dynamical systems}

\subsubsection{Simulation}

\newpage

\begin{figure}[h!]
    \centering
    \includegraphics[width=1\linewidth]{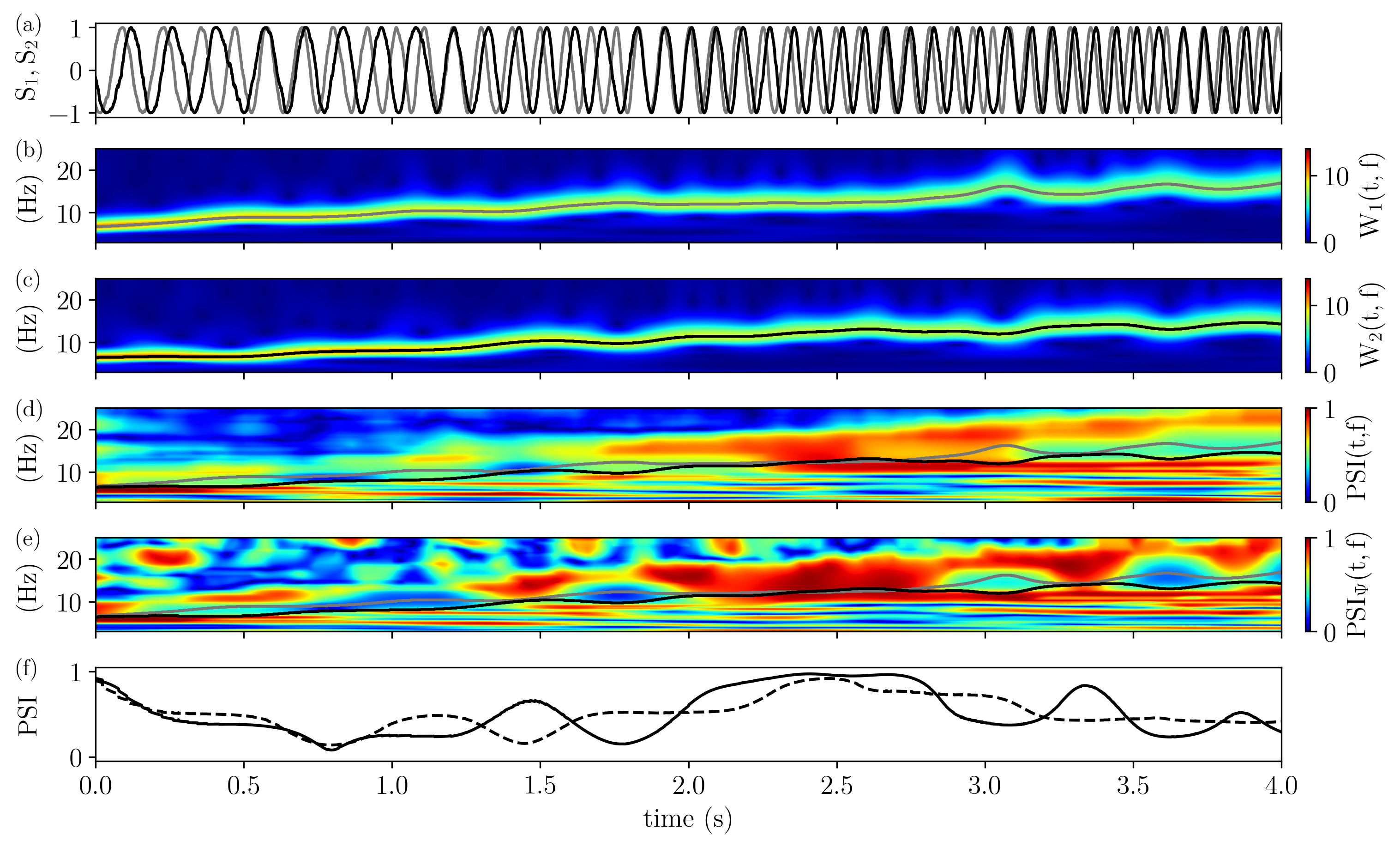}
    \caption{\textbf{System of two coupled oscillators with a linear frequency evolution (chirp).}
    (a) Simulated signal following Eq.~(\ref{eq:S_delta_noise}). Gray: $\cos(\varphi_1)$. Black: $\cos(\varphi_2)$. Frequencies: $\omega_1/2\pi=\qty{5}{\hertz}$, $\omega_2/2\pi=\qty{4}{\hertz}$.
    Coupling strength: $B_1/2\pi=B_2/2\pi=\qty{0.5}{\hertz}$
    (b) $|W_1(t,f)|$, $n_0=6$. Gray line: local frequency maxima $f_1^*(t)$.
    (c) $|W_2(t,f)|$, $n_0=6$. Black line: local frequency maxima $f_2^*(t)$.
    (d) $\Upsilon(t,f)$ computed with a constant window $\Delta T = 1s$. Gray: $f_1^*(t)$. Black: $f_2^*(t)$.
    (e) $\Upsilon_\Psi(t,f)$ computed with a variable window $\Delta T_\Psi = n_0/f^*$ (Eq.~\ref{eq:PSI_CWT_complex}), $n_0=6$. Gray: $f_1^*(t)$. Black: $f_2^*(t)$.
    (f) $\Upsilon_\Psi (t,f^*_{12}(t))$, where $f^*_{12}(t) = (f_1^*(t) + f_2^*(t))/2$, computed using a constant window $\Delta T = 1s$ (dashed line) and a variable window $\Delta T_\Psi = n_0/f^*$ (plain line).}
    \label{fig:SUP_PSI_on_sim}
\end{figure}

Fig.~\ref{fig:SUP_PSI_on_sim}(d) displays a time frequency PSI: $\Upsilon$, where temporal averaging is performed over a constant time window $\Delta T = 1s$. This basic PSI computation does not take into account the dynamics of the system and loses its precision over time when the overall frequencies of the two oscillators increase. This figure illustrates the relevance of computing the PSI over a real-time adaptive wavelet window $\Delta T_\Psi$ (Fig.~\ref{fig:SUP_PSI_on_sim}(e)) using Eq.~\ref{eq:PSI_CWT_complex}, where $\Delta T _ \Psi= [\Delta T_ \Psi (f^*_1) + \Delta T_ \Psi (f^*_2)] / 2$.

\clearpage

\subsubsection{Generalization of PSI computation}

The expression of the time frequency PSI provided in Eq.~\ref{eq:PSI_CWT_complex} uses the hypothesis that the two oscillators have a common frequency $f_1 = f_2$, and correspond to the case $n=m=1$. However, since the frequency of each oscillator may drift or even jump to a distant value with time, as can be noticed in experimental spectrograms (Fig.~\ref{fig:resync}(b-c)), the hypothesis $n$, $m$ constant, with $n=m = 1$ is no longer relevant. To compute the synchronization of oscillatory components with different or non constant frequencies, we need to reformulate Eq.~\ref{eq:PSI_CWT_complex} \cite{pikovsky_synchronization_2002}. 

If we consider two oscillators with different unperturbed (natural) frequencies: 
\begin{equation}
    \varphi_1 = \omega_1 t \; \; , \; \; \varphi_2 = \omega_2 t ,
\end{equation}
with a rational ratio of the frequencies: 
\begin{equation}
    \frac{\omega_1}{\omega_2} \simeq \frac{m}{n} ,
\end{equation}
corresponding to a nearly resonance of the two oscillators, one can generalize the dynamical equations for the phases $\varphi_1$ and $\varphi_2$ as: 
\begin{equation}
\left\{ 
    \begin{array}{l}
        \dot{\varphi_1} = \omega_1 + \varepsilon q_1(n\varphi_1 - m\varphi_2) \\
        \dot{\varphi_2} = \omega_2 + \varepsilon q_2 (m\varphi_2 - n\varphi_1)
    \end{array} \; . \label{eq:coupled_phase_rev}
\right.
\end{equation}
where $\varepsilon$ is a constant coupling strength and the $q_{1,2}$ functions can be written as Fourier series
\begin{equation}
\left\{ 
    \begin{array}{l}
        q_1(n\varphi_1 - m\varphi_2) = \sum_j a^{(1)}_{nj,-mj} e^{i \; (j (n\varphi_1 -m \varphi_2))} \\
        q_2(m\varphi_2 - n\varphi_1) = \sum_j a^{(2)}_{mj,-nj} e^{i \; (j (m \varphi_2 - n\varphi_1 ))}
    \end{array} \; . \label{eq:coupled_phase_qij}
\right.  \; ,
\end{equation}
$i = (-1)^{1/2}$, $j$ is an positive integer index for the sum, the $a^{(1,2)}$ are Fourier coefficients for these decompositions. Here the $nj, -mj$ factors are selected as corresponding to resonant cases. 

For the difference of phase of the two oscillators: $\psi = n\phi_1 -m \phi_2$ we obtain: 
\begin{equation}
    \frac{d\psi}{dt} = -\nu + \varepsilon q(\psi) ,
\end{equation}
where $\nu = m\omega_2 -n \omega_1$ and $q(\psi) = n q_1(\psi) -m q_2(\psi)$ .

We observe that when the two oscillators have natural frequencies in a rational fraction, near resonance, the dynamics of the phase difference follows slower dynamics which oscillate at the difference of pulsations rescaled by the coefficients $m$ and $n$ of the rational fraction. 

\begin{figure}[!h]
    \centering
    \includegraphics[width=1\linewidth]{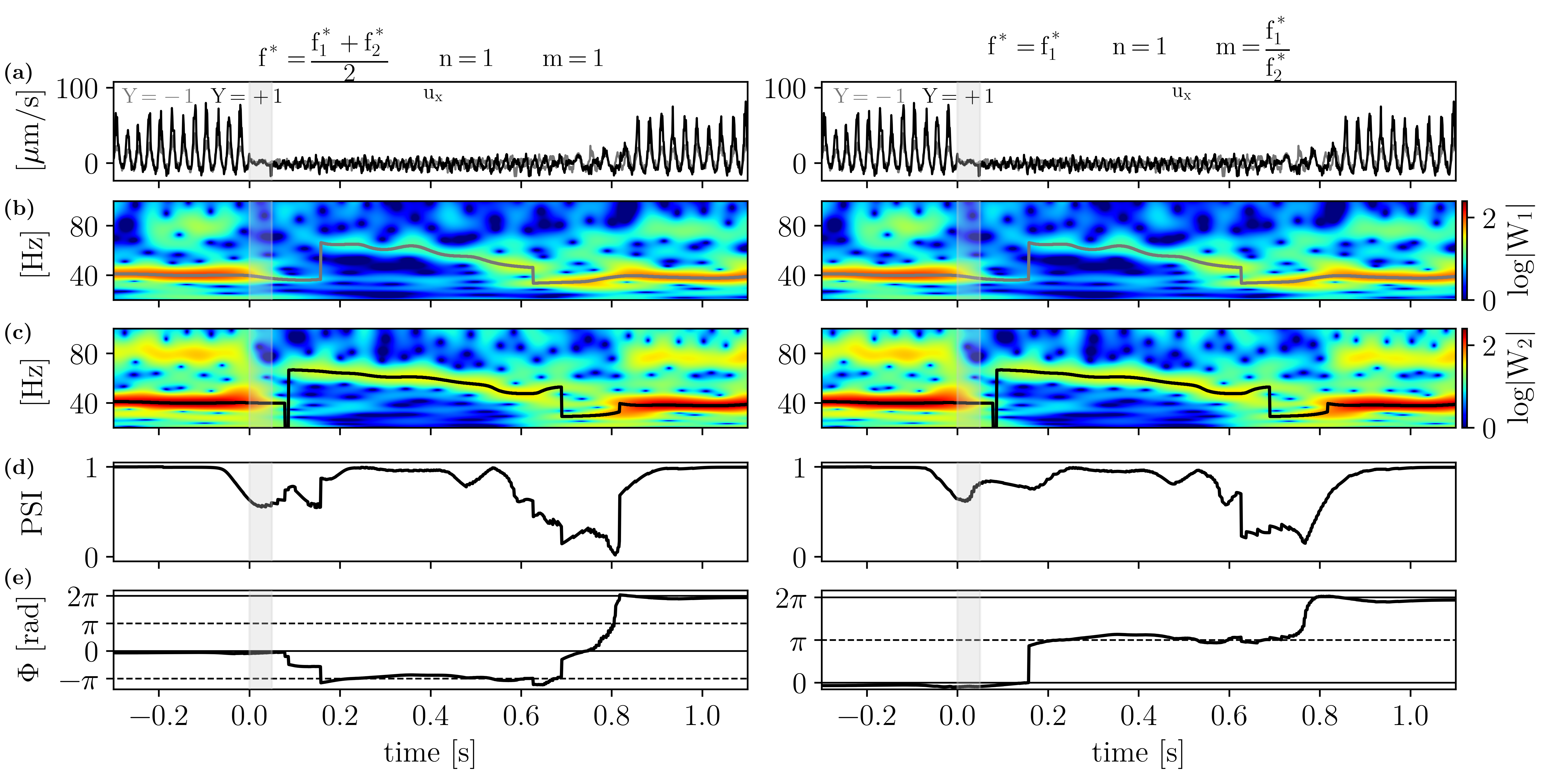}
    \caption{\textbf{Generalization of $\widetilde{\Upsilon}_\Psi$ computation with the hypothesis $n \ne m$}
    The left column is identical to that of Fig.\ref{fig:resync} where $n=m=1$ and the local frequency used for computing the width of the averaging window $f^* = (f_1^* + f_2^*)/2$, the right column uses $n=1, m=(f_1^*/f_2^*)$ and does not average the local frequency, selecting the local frequency of oscillator 1: $f^* = f_1$.
    (a) Averaged fluid velocity signals along the $x$-axis, extracted from regions on each side of the alga: $Y=-1$ (gray - oscillator 1) and $Y=+1$ (black - oscillator 2).
    (b-c) Time-frequency spectrograms $\log |W_Y(t,f)|$ (Morlet wavelet with $n_0=12$), for $Y=-1$ (b) and $Y=+1$ (c). Frequency ridges $f_{-1}^*(t)$ (gray) and $f_{+1}^*(t)$ (black) are also plotted.
    (d) Phase Synchronization Index, $|\widetilde{\Upsilon}_\Psi| (t, f^*_{\pm 1}(t))$.
    (e) Phase $\Phi(t)$ obtained from the complex PSI $\widetilde{\Upsilon}_\Psi$.
In all panels, the shaded gray region marks the duration of the photoshock.}
    \label{fig:new_PSI_generalized}
\end{figure}

Hence, when we pick two oscillatory dynamics from the time-frequency decomposition to compute their PSI, we need to rescale their respective phases accordingly. If we compare two maxima frequency lines (from two different oscillatory modes (fundamental and first harmonic) or from the same fundamental mode of two oscillators), we need to use the local frequencies in correct position: 

\begin{equation}
\widetilde{\Upsilon}_{\Psi}^{(g)} (t,f) = \left< \frac{W_1(t,f)^n \overline{W_2}(t,f)^m}{|W_1(t,f)|^n|W_2(t,f)|^m} \right>_{\Delta T _{\Psi}} = \Upsilon_\Psi \; e^{i\Phi} \; \label{eq:PSI_CWT_complex_generalized}.
\end{equation}
Here the temporal averaging $\left<\cdot\right>_{\Delta T _{\Psi}}$ is performed over the wavelet window $\left[t-\frac{\Delta T _{\Psi}}{2},t+\frac{\Delta T _{\Psi}}{2}\right]$, with $\Delta T _{\Psi} \propto n_0 / f^*_{1}$, 
$f^*_{1}$ being the local frequency of oscillator 1. This computation is illustrated in Fig.~\ref{fig:new_PSI_generalized}, where we note that the correction is efficient in the intermediate regime (just after the photoshock) where the two oscillators undergo a drastic transition in frequency (modification of the swimming mode). In the plots of both the PSI modulus and its phase, the oscillator 1 has already jumped to the other mode, whereas the oscillator 2 stays for roughly \qty{100}{\milli\second} closer to its initial mode. The $-\pi /2$ jump is clearly an artifact of the computation, assuming that $n=m=1$, whereas it is estimated at $n=1, m= f_1^*/f_2^* \sim 1.8$. On the right panel of Fig.~\ref{fig:new_PSI_generalized}, one can observe that the artifact is corrected, and the $-\pi/2$ plateaus vanishes while leaving the in-phase to anti-phase transition present. 

\clearpage

\subsection{Breaststroke frequency of selected algae}

\begin{figure}[h!]
    \centering
    \includegraphics[width=\linewidth]{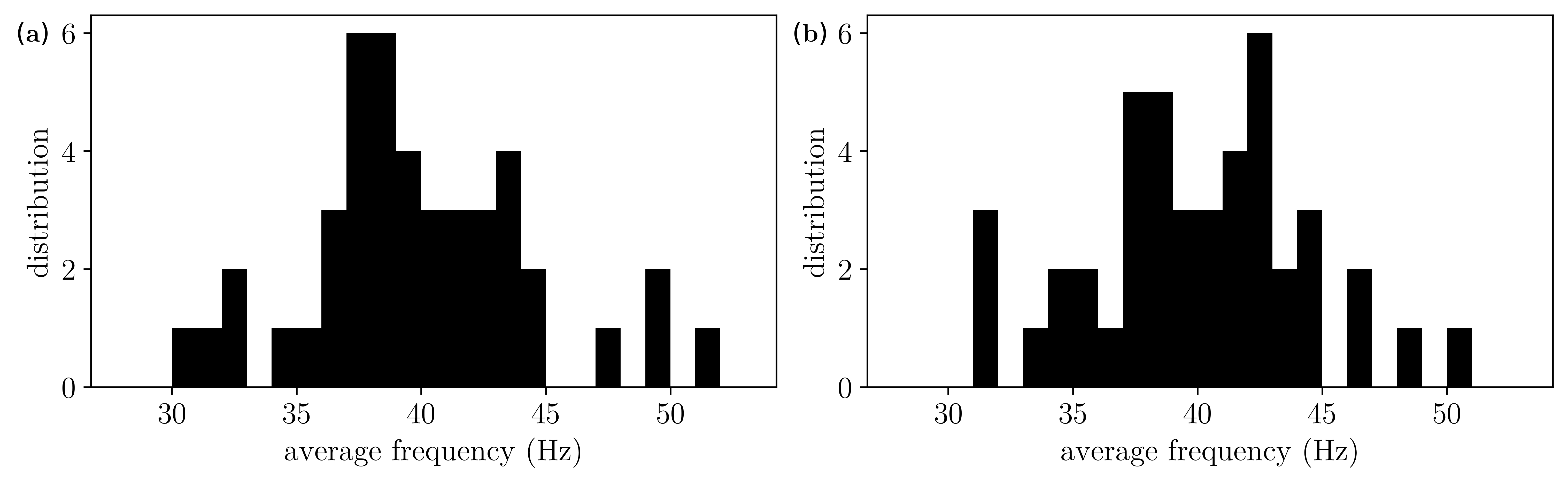}
    \caption{\textbf{Distribution of the mean beating frequency observed in breaststroke swimming ($\qty{1}{\hertz}$ bins).} The mean of the Wavelet transform Modulus of the $u_x$ signal (Y=0), is computed on breaststroke swimming regions and the frequency associated to the maxima is taken. Total of 44 frequencies, associated to 44 photoshock runs on 13 different cells.
    (a) Average frequency distribution before the photoshock (between $t=-\qty{1000}{\milli\second}$ and $t=-\qty{50}{\milli\second}$ [Fig.~\ref{fig:freq_maxima}]).
    (b) Average frequency distribution after the photoshock (between $t=\qty{2000}{\milli\second}$ and $t=\qty{3000}{\milli\second}$ [Fig.~\ref{fig:freq_maxima}]).
    }
    \label{fig:SUP_avg_freq}
\end{figure}

\clearpage

\subsection{Dynamics over multiple photoshocks}

\end{document}